\documentclass[aps,showpacs,preprintnumbers,amsmath, amssymb]{revtex4}

\usepackage{float}
\usepackage{graphics,epsfig}
\usepackage{graphicx}
\usepackage{dcolumn}
\usepackage{bm}

\begin{document}
\baselineskip=0.8 cm
\title{{\bf Holographic superconductors in 4D Einstein-Gauss-Bonnet gravity}}

\author{Xiongying Qiao$^{1}$, Liang OuYang$^{1}$, Dong Wang$^{1}$, Qiyuan Pan$^{1,2}$\footnote{panqiyuan@hunnu.edu.cn}, and Jiliang Jing$^{1,2}$\footnote{jljing@hunnu.edu.cn}}
\affiliation{$^{1}$Key Laboratory of Low Dimensional Quantum Structures and Quantum Control of Ministry of Education, Synergetic Innovation Center for Quantum Effects and Applications, and Department of Physics, Hunan Normal University, Changsha, Hunan
410081, China} \affiliation{$^{2}$Center for Gravitation and Cosmology, College of Physical Science and Technology, Yangzhou University, Yangzhou 225009, China}

\vspace*{0.2cm}
\begin{abstract}
\baselineskip=0.6 cm
\begin{center}
{\bf Abstract}
\end{center}

We investigate the neutral AdS black-hole solution in the consistent $D\rightarrow4$ Einstein-Gauss-Bonnet gravity proposed in [K. Aoki, M.A. Gorji, and S. Mukohyama, Phys. Lett. B {\bf 810}, 135843 (2020)] and construct the gravity duals of ($2+1$)-dimensional superconductors with Gauss-Bonnet corrections in the probe limit. We find that the curvature correction has a more subtle effect on the scalar condensates in the s-wave superconductor in ($2+1$)-dimensions, which is different from the finding in the higher-dimensional superconductors that the higher curvature correction makes the scalar hair more difficult to be developed in the full parameter space. However, in the p-wave case, we observe that the higher curvature correction always makes it harder for the vector condensates to form in various dimensions. Moreover, we note that the higher curvature correction results in the larger deviation from the expected relation in the gap frequency $\omega_g/T_c\approx 8$ in both ($2+1$)-dimensional s-wave and p-wave models.

\end{abstract}


\maketitle
\newpage
\vspace*{0.2cm}

\section{Introduction}

In condensed matter physics, the mechanism leading to superconductivity with high critical temperature, which can not be described by the conventional Bardeen-Cooper-Schrieffer (BCS) theory \cite{BCS}, is still unclear. However, recently developed holographic superconductor models, have been used to give some insights into the pairing mechanism in the high-temperature superconductor systems \cite{HartnollRev,HerzogRev,HorowitzRev,CaiRev}. According to the anti-de Sitter/conformal field theory (AdS/CFT) correspondence, which relates strongly coupled systems to weakly coupled systems \cite{Maldacena}, Gubser presented the spontaneous symmetry breaking of the $U(1)$ symmetry for an Abelian Higgs model coupled to the gravity theory with a negative cosmological constant \cite{GubserPRD78}, and Hartnoll \emph{et al.} reproduced the properties of a ($2+1$)-dimensional s-wave superconductor in the ($3+1$)-dimensional holographic superconductor model based on the framework of usual Maxwell electrodynamics \cite{HartnollPRL101}. To go a step further, the p-wave holographic superconductivity was realized by introducing an $SU(2)$ Yang-Mills field into the bulk \cite{GubserPufu} while the d-wave holographic superconductor was constructed by introducing a charged massive spin two field propagating in the bulk \cite{DWaveChen,DWaveBenini}. Especially, introducing a charged vector field into an Einstein-Maxwell theory \cite{CaiPWave-1}, Cai \emph{et al.} developed a
novel p-wave holographic superconductor model which is a generalization of the $SU(2)$ model with a general mass
and gyromagnetic ratio \cite{CaiPWave-2}. Considering the application of the Mermin-Wagner theorem to the holographic superconductors, Gregory \emph{et al.} constructed holographic superconductors in the five-dimensional Einstein-Gauss-Bonnet gravity in the probe limit where the backreaction of matter fields on the spacetime metric is neglected \cite{Gregory}. It was observed that the higher curvature correction makes the scalar condensates harder to form and causes the behavior of the claimed universal ratio $\omega/T_c\approx8$ \cite{HorowitzPRD78} unstable. Note that this conclusion still holds in higher dimensions \cite{Pan-Wang} and beyond the probe limit \cite{BarclayGregory,Brihaye,Gregory2011}. Other generalized investigations based on the effects of the curvature correction on the holographic dual models can be found, for example, in Refs. \cite{Ge-Wang,KannoGB,Gangopadhyay2012,GhoraiGangopadhyay,
SheykhiSalahiMontakhab,SalahiSheykhiMontakhab,LiFuNie,CHNam,ParaiEPJC2020,CaiPWaveGB,
Pan-Jing-Wang-Soliton,LiCaiZhang,NieZeng,LuWuNPB2016,GBSuperfluid,MohammadiEPJC2019}.

All the studies mentioned above concerning the holographic dual models with the curvature correction are based on the Einstein-Gauss-Bonnet gravity in dimensions $D\geq5$, where we find that the higher curvature corrections make it harder for the scalar \cite{Gregory,Pan-Wang,BarclayGregory,Brihaye,Gregory2011,Ge-Wang,KannoGB,
Gangopadhyay2012,GhoraiGangopadhyay,SheykhiSalahiMontakhab,SalahiSheykhiMontakhab,LiFuNie,CHNam,ParaiEPJC2020} or vector \cite{CaiPWaveGB,Pan-Jing-Wang-Soliton,LiCaiZhang,NieZeng,LuWuNPB2016,GBSuperfluid,MohammadiEPJC2019} hair to form. As pointed out by Gregory \emph{et al.} in \cite{Gregory}, \textit{one can expect this tendency to be the same even in ($2+1$)-dimensions, however, it remains obscure to what extent this suppression affects the physics of holographic superconductors in ($2+1$)-dimensions}. It now seems likely that we can investigate the influence of the curvature correction on the ($2+1$)-dimensional superconductors because of the recently introduced Einstein-Gauss-Bonnet gravity in four dimensions \cite{GlavanLin}. Rescaling the Gauss-Bonnet coupling constant $\alpha\rightarrow\alpha/(D-4)$ and taking the limit $D\rightarrow4$ \cite{GlavanLin}, Glavan and Lin formulated the novel four-dimensional (4D) Einstein-Gauss-Bonnet gravity and reported several appealing new predictions of this theory, including the singularity resolution for spherically symmetric solutions. Extending the investigation to the AdS space, the authors obtained the charged AdS black-hole solution in the 4D Einstein-Gauss-Bonnet Gravity \cite{Fernandes} and Einstein-Lovelock gravity \cite{KonoplyaZhidenko}. Interestingly, the solution of the same form has been presented in the conformal anomaly inspired gravity \cite{CaiCaoOhta}. Along this line, the Einstein-Gauss-Bonnet gravity in the 4D spacetime has been explored extensively on various aspects, including the exact solutions \cite{CasalinoCRV,KumarGhoshMaharaj,SinghGM,DonevaYazadjiev,JusufiBG,GeSin,Liu14267,Yang14468,MaLu}, the quasinormal modes and stability \cite{KonoplyaZinhailoZhidenko,Churilova,Mishra,LiWY,ZhangZLG,AragonBGV,MalafarinaTD,Cuyubamba09025,LiuNZ,Devi14935}, the observable shadows \cite{GuoLi,WeiLiu07769,ZhangWeiLiu,Heydari-Fard,RayimbaevATA,ZengZZ}, the geodesics and gravitational lensing \cite{LiuZW,KumaraRHAA,IslamKG,JinGL,Kumar12970}, and the thermodynamics and cosmic censorship conjecture \cite{HegdeKA,SinghS,ZhangLG,Mansoori,WeiL14275,PanahJafarzade,YangWCYW,Ying}.

However, the original claim of Ref. \cite{GlavanLin} is clearly in contradiction with the Lovelock theorem, and some subtleties and criticisms on the $D\rightarrow4$ limit were revealed in \cite{Ai2020,Mahapatra2020,Shu09339,TianZhu,ArrecheaDJ}, concluding that there is no pure four-dimensional Einstein-Gauss-Bonnet gravity \cite{GursesST}. In Refs. \cite{LuPang,HennigarKMP}, the authors proposed the ``regularized" versions of the 4D Einstein-Gauss-Bonnet gravity, i.e., the scalar-tensor description of the $D\rightarrow4$ limit of the Einstein-Gauss-Bonnet gravity, which can be described effectively by a particular subclass of the Horndeski theory. Other investigations based on the regularized 4D Einstein-Gauss-Bonnet gravity can be found, for example, in Refs. \cite{Fernandes08362,LuMao}. However, Kobayashi pointed out that the regularized 4D Einstein-Gauss-Bonnet gravity contains an additional dynamical scalar field but the scalar field lacks the quadratic kinetic term, which leads to the infinite strong coupling problem \cite{Kobayashi}. In Ref. \cite{BonifacioHJ}, this issue was also discussed by Bonifacio \emph{et al.} in detail. Fortunately, using the Arnowitt-Deser-Misner (ADM) decomposition, Aoki \emph{et al.} proposed a novel four-dimensional theory that serves as a consistent realization of the $D\rightarrow4$ limit of the Einstein-Gauss-Bonnet gravity with two dynamical degrees of freedom by breaking the temporal diffeomorphism invariance \cite{AokiGMPLB}. In particular, it was found that the Friedmann-Lema\^{i}tre-Robertson-Walker (FLRW) and black hole solutions obtained in \cite{GlavanLin} are solutions of this well-defined theory, which has been checked explicitly in \cite{AokiGMJCAP}. In Ref. \cite{Konoplya02248}, Konoplya \emph{et al.} investigated the grey-body factor for Dirac, electromagnetic and gravitational fields, and estimated the intensity of Hawking radiation and lifetime for asymptotically flat black holes in this consistent theory. More recently, the authors of Ref. \cite{AokiGMM} studied the inflationary gravitational waves in the consistent $D\rightarrow4$ Einstein-Gauss-Bonnet gravity and found a new attractor regime which they called the Gauss-Bonnet attractor as the dominant contribution coming from the Gauss-Bonnet term.

In this work, we will use the consistent theory of the $D\rightarrow4$ Einstein-Gauss-Bonnet gravity \cite{AokiGMPLB} to give the 4D Gauss-Bonnet-AdS black hole solution, which is proved out to be the solution obtained via the naive $D\rightarrow4$ limit of the higher-dimensional theory (by setting the charge $Q=0$) \cite{Fernandes,KonoplyaZhidenko} as well as the scalar-tensor theories with the Gauss-Bonnet term \cite{LuPang,HennigarKMP}. Considering the increasing interest in study of the 4D Einstein-Gauss-Bonnet gravity, we are going to examine the influence of the curvature correction on the ($2+1$)-dimensional superconductor models which has not been studied as far as we know. We will give a more complete picture of how the curvature correction affects the condensate for the scalar/vector operator and the conductivity by introducing the s-wave and p-wave holographic superconductors with Gauss-Bonnet corrections in four dimensions, and compare the result with that given in the higher-dimensional Gauss-Bonnet superconductors. In order to avoid the complex computation and extract the main physics, we will concentrate on the probe limit neglecting backreaction of the spacetime.

The plan of the work is the following. In Sec. II we will present the 4D neutral AdS black-hole solution in the Einstein-Gauss-Bonnet gravity by using the consistent $D\rightarrow4$ Einstein-Gauss-Bonnet gravity proposed in \cite{AokiGMPLB}, which can also be obtained from Ref. \cite{KonoplyaZhidenko} by setting the charge $Q=0$. In Sec. III we will construct the s-wave holographic superconductors in the 4D Einstein-Gauss-Bonnet gravity and investigate the effect of the curvature correction on the ($2+1$)-dimensional superconductors. In Sec. IV we will explore the p-wave cases. We will conclude in the last section with our main results.

\section{4D Gauss-Bonnet-AdS black holes}
For the consistent theory of the four-dimensional Einstein-Gauss-Bonnet gravity, we can write the metric in the ADM formalism as
\begin{equation}\label{metric}
ds^2=g_{\mu\nu} dx^{\mu} dx^{\nu}=-N^{2}dt^{2}+\gamma_{ij}(dx^{i}+N^{i}dt)(dx^{j}+N^{j}dt),
\end{equation}
with the lapse function $N$, shift vector $N^i$ and spatial metric $\gamma_{ij}$. And we can express the gravitational action as \cite{AokiGMPLB,AokiGMJCAP,AokiGMM}
\begin{eqnarray}\label{4DEGBAction}
S&=& \int dt d^3x N\sqrt{\gamma}\mathcal{L}^{\rm 4D}_{\rm EGB},
\end{eqnarray}
with the Lagrangian density
\begin{eqnarray}\label{LD-EGB}
\mathcal{L}^{\rm 4D}_{\rm EGB} &=& \frac{M_{\rm Pl}^2}{2}
\left \{2R+\frac{6}{L^{2}}- \mathcal{M} + \frac{\alpha}{2}
\left [8R^2 -4 R\mathcal{M} -\mathcal{M}^2
- \frac{8}{3} \left(8R_{ij}R^{ij}-4R_{ij}\mathcal{M}^{ij}
-\mathcal{M}_{ij}\mathcal{M}^{ij}\right) \right ] \right \}\,, \nonumber
\end{eqnarray}
where $M^{2}_{\rm Pl}=(8\pi G)^{-1}$ is the reduced Planck mass. $R$ and $R_{ij}$ are respectively the Ricci scalar and Ricci tensor of the spatial metric, and
\begin{eqnarray}\label{M}
\mathcal{M}_{ij} \equiv R_{ij}+\mathcal{K}^k{}_{k}
\mathcal{K}_{ij}-\mathcal{K}_{ik}\mathcal{K}^k{}_{j},
\hspace{1cm} \mathcal{M} \equiv \mathcal{M}^i{}_{i} \,,
\end{eqnarray}
with
\begin{eqnarray}\label{Kij}
\mathcal{K}_{ij} \equiv \frac{1}{2N}\left [\dot{\gamma}_{ij}-2D_{(i}N_{j)}-\gamma_{ij}D^2 \lambda_{\rm
GF} \right ],
\end{eqnarray}
where a dot denotes the derivative with respect to the time $t$, and $D_{i}$ represents the covariant derivative
compatible with the spatial metric.

Considering the four-dimensional topological black hole, we take the metric ansatz
\begin{equation}\label{metric-BG}
N=e^{A(r)} \,, \hspace{1cm} N^i = 0 \,, \hspace{1cm} \gamma_{ij}=\text{diag}(e^{2B(r)},r^2,r^2H^{(k)})
\end{equation}
with
\begin{eqnarray}\label{H4}
H^{(k)}=\left\{
\begin{array}{ll} \sinh^2\theta\;,\;\;\;\;\; k=-1;\\
1,\qquad \qquad k=0;\\
\sin^2\theta \;,\;\;\;\;\;\;\; k=1.
\end{array}
\right.
\end{eqnarray}
For the symmetric static backgrounds, we can set the Lagrange multiplier $\lambda_{\rm GF}$ to zero in practice and (\ref{Kij}) reduces to the standard extrinsic curvature \cite{AokiGMJCAP,AokiGMM}. Thus, we get the simple form of the action
\begin{equation}
\label{EffAction}
S=\frac{\Sigma_k}{8\pi G}\int dt\ dr e^{A+B}\left [r^{3}\varphi\left(1+\alpha\varphi\right ) +\frac{r^{3}}{L^2}\right ]',
\end{equation}
where $\Sigma_k$ represents the volume, $\varphi=r^{-2}(k-e^{-2B})$ is the function of $r$ only, and the prime denotes the derivative with respect to $r$. Thus, from this action we have  \cite{Cai-2002}
\begin{eqnarray}
e^{A+B}=1,~~~\varphi(1+\alpha\varphi)+\frac{1}{L^2}=\frac{8\pi G M}{\Sigma_k r^{3}}.
\end{eqnarray}
Obviously, we can get the exact solution
\begin{equation}
\label{solution}
e^{2A} =e^{-2B}=k +\frac{r^2}{2\alpha}\left(1\pm
\sqrt{1+\frac{32\pi\alpha GM}{\Sigma_k r^{3}}-\frac{4\alpha}{L^2}}\right),
\end{equation}
which coincides with the black-hole solution obtained via the naive $D\rightarrow4$ limit of the higher-dimensional theory (by setting the charge $Q=0$) \cite{Fernandes,KonoplyaZhidenko} as well as the regularized 4D Einstein-Gauss-Bonnet gravity proposed in Refs. \cite{LuPang,HennigarKMP}. Since we can recover the Schwarzschild-AdS black hole in the limit $\alpha\rightarrow0$, we are more interested in the physically interesting solution with the minus sign.

In order to construct a superconductor dual to an AdS black hole configuration in the probe
limit, in this work we consider the background of the 4D planar Gauss-Bonnet-AdS black hole
\begin{equation}
\label{planarsolution}
f=e^{2A} =e^{-2B}=\frac{r^2}{2\alpha}\left[1-
\sqrt{1-\frac{4\alpha}{L^2}\left(1-\frac{r^{3}_{+}}{r^{3}}\right)}\right],
\end{equation}
where the black hole horizon $r_{+}$ is related to the mass $M$ through
$r_{+}^{3}=8\pi GML^{2}/\Sigma_{0}$ with the volume of the Ricci flat space $\Sigma_{0}$. It should be noted that in the
asymptotic region, i.e., $r\rightarrow\infty$, we find $f(r)\sim r^2\left(1-\sqrt{1-4\alpha/L^2}\right)/\left(2\alpha\right)$,
which means that, just like the cases $D\geq5$, we can define the effective asymptotic AdS scale by \cite{Cai-2002}
\begin{eqnarray}\label{LeffAdS}
L^2_{\rm eff}=\frac{2\alpha}{1-\sqrt{1-\frac{4\alpha}{L^2}}},
\end{eqnarray}
where the so-called Chern-Simons limit $\alpha=L^{2}/4$ \cite{CrisostomoTZ} is the upper bound of the Gauss-Bonnet parameter. For simplicity, in the following we will consider the Gauss-Bonnet parameter in the range $-L^{2}/10\leq\alpha\leq L^{2}/4$. We can easily obtain the Hawking temperature of this 4D Gauss-Bonnet-AdS black hole
\begin{eqnarray}
\label{Hawking temperature}
T=\frac{3r_{+}}{4\pi L^{2}},
\end{eqnarray}
which will be interpreted as the temperature of the CFT. For convenience, we will scale $L=1$ in the following calculation.

\section{s-Wave holographic superconductors}

We want to know the influence of the curvature correction on the ($2+1$)-dimensional superconductors since we have obtained the 4D Gauss-Bonnet-AdS black hole by using the consistent theory proposed in \cite{AokiGMPLB}. In this section, we first introduce the s-wave holographic superconductor models in the 4D Gauss-Bonnet-AdS black hole.

\subsection{Condensates of the scalar field}

We consider a Maxwell field and a charged complex scalar field coupled via the action
\begin{eqnarray}\label{sWaveSystem}
S=\int d^{4}x\sqrt{-g}\left[
-\frac{1}{4}F_{\mu\nu}F^{\mu\nu}-g^{\mu\nu}(\nabla_{\mu}\psi-iqA_{\mu}\psi)(\nabla_{\nu}\psi-iqA_{\nu}\psi)^{\ast}
-m^2|\psi|^2 \right].
\end{eqnarray}
Adopting the ansatz for the matter fields $\psi=|\psi|$, $A_{t}=\phi$ where $\psi$, $\phi$
are both real functions of $r$ only, we arrive at the equations of
motion
\begin{eqnarray}
&&\psi^{\prime\prime}+\left(
\frac{2}{r}+\frac{f^\prime}{f}\right)\psi^\prime
+\left(\frac{q^{2}\phi^2}{f^2}-\frac{m^2}{f}\right)\psi=0,
\label{BHPsi}
\end{eqnarray}
\begin{eqnarray}
\phi^{\prime\prime}+\frac{2}{r}\phi^\prime-\frac{2q^{2}\psi^{2}}{f}\phi=0, \label{BHPhi}
\end{eqnarray}
where the prime denotes differentiation in $r$.

To get the solutions in the superconducting phase, i.e., $\psi(r)\neq0$, we have to count on the appropriate boundary conditions. At the horizon $r=r_{+}$, the regularity leads to the boundary conditions $\psi(r_{+})=f^\prime(r_{+})\psi^\prime(r_{+})/m^{2}$ and $\phi(r_{+})=0$. Near the AdS boundary $r\rightarrow\infty$, the asymptotic behaviors of the solutions are
\begin{eqnarray}
\psi=\frac{\psi_{-}}{r^{\lambda_{-}}}+\frac{\psi_{+}}{r^{\lambda_{+}}}\,,\hspace{0.5cm}
\phi=\mu-\frac{\rho}{r}\,, \label{infinity}
\end{eqnarray}
with the characteristic exponents
\begin{eqnarray}
\lambda_\pm=\frac{1}{2}\left(3\pm\sqrt{9+4m^{2}L_{\rm eff}^2}\right), \label{characteristicexponent}
\end{eqnarray}
where $\mu$ and $\rho$ are interpreted as the chemical potential and charge density in the dual field theory, respectively. Considering the stability of the scalar field, we observe that the mass should be above the Breitenlohner-Freedman bound $m^{2}_{BF}=-9/(4L_{\rm eff}^{2})$ \cite{Breitenloher}, which depends on the Gauss-Bonnet parameter $\alpha$. From the AdS/CFT correspondence, provided $\lambda_{-}$ is larger than the unitarity bound, both $\psi_{-}$ and $\psi_{+}$ can be normalizable and correspond to the vacuum expectation values $\langle O_{-}\rangle=\psi_{-}$, $\langle O_{+}\rangle=\psi_{+}$ of an operator $O$ dual to the scalar field, respectively. We will impose boundary condition that either $\psi_{+}$ or $\psi_{-}$ vanishes since imposing boundary conditions in which both $\psi_{-}$ and $\psi_{+}$ are non-zero makes the
asymptotic AdS theory unstable \cite{HartnollPRL101}.

It is interesting to note that, from Eqs. (\ref{BHPsi}) and (\ref{BHPhi}), we can get the useful
scaling symmetries and the transformation of the relevant quantities
\begin{eqnarray}
&&r\rightarrow \beta r,~~~(t,x,y)\rightarrow
\beta^{-1}(t,x,y),~~~q\rightarrow q,~~~\psi\rightarrow\psi,~~~\phi\rightarrow
\beta\phi,
\nonumber \\
&&(T,\mu)\rightarrow
\beta(T,\mu),~~~\rho\rightarrow\beta^{2}\rho,~~~\psi_{\pm}\rightarrow
\beta^{\lambda_{\pm}}\psi_{\pm},
\end{eqnarray}
where $\beta$ is a real positive number. We will make use of these properties to set $r_{+}=1$ and $q=1$ when
performing numerical calculations.

\subsubsection{Scalar operator $O_{+}$}

Using the shooting method \cite{HartnollPRL101}, we can solve numerically the equations of motion (\ref{BHPsi}) and (\ref{BHPhi}) by doing an integration from the horizon out to the AdS boundary. From the expression (\ref{characteristicexponent}), we can choose the mass of the scalar field by selecting the values of $m^{2}L^{2}$ or $m^{2}L_{\rm eff}^{2}$, due to the presence of the Gauss-Bonnet parameter, just as in the cases of $D\geq5$ \cite{Gregory,Pan-Wang}. Obviously, we have to reinvestigate the issue related to the different choices of the mass
of the scalar field for the ($2+1$)-dimensional Gauss-Bonnet superconductors.

\begin{figure}[ht]
\includegraphics[scale=0.65]{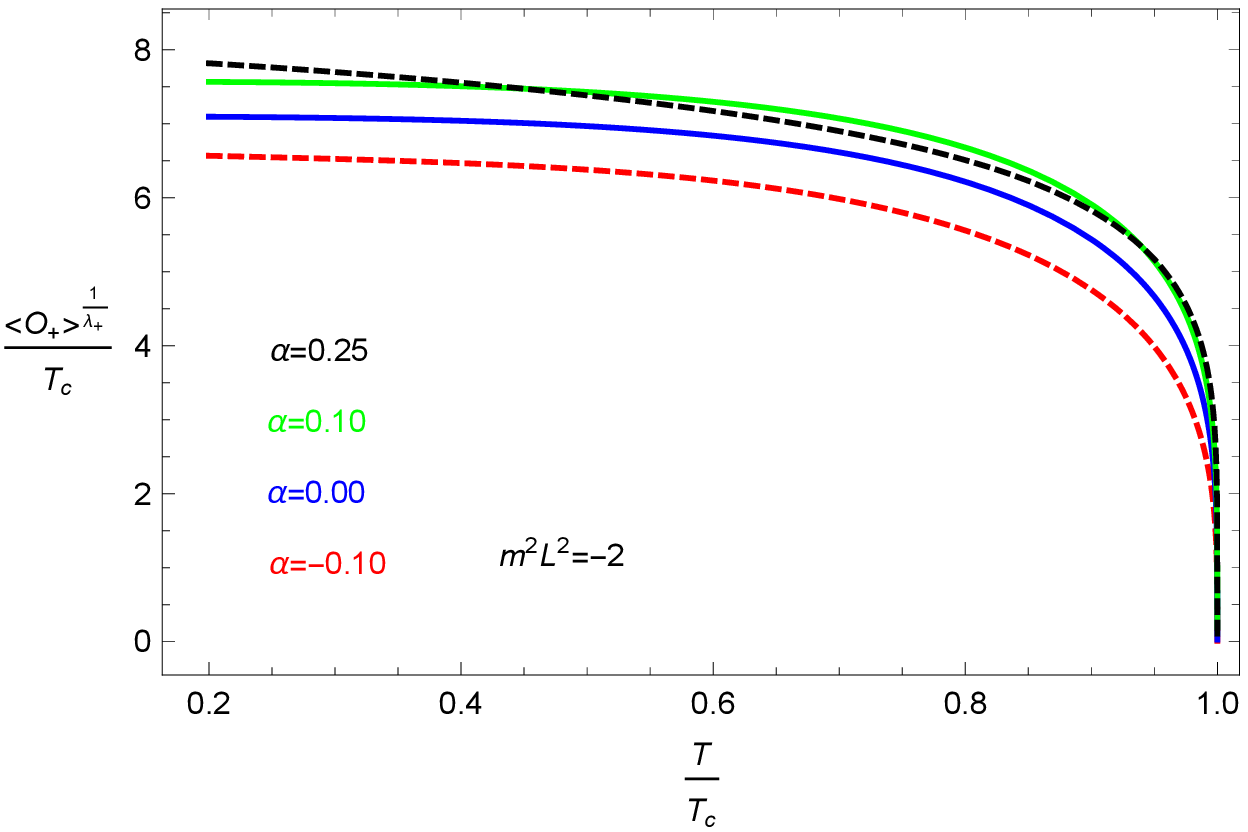}\vspace{0.0cm}
\includegraphics[scale=0.65]{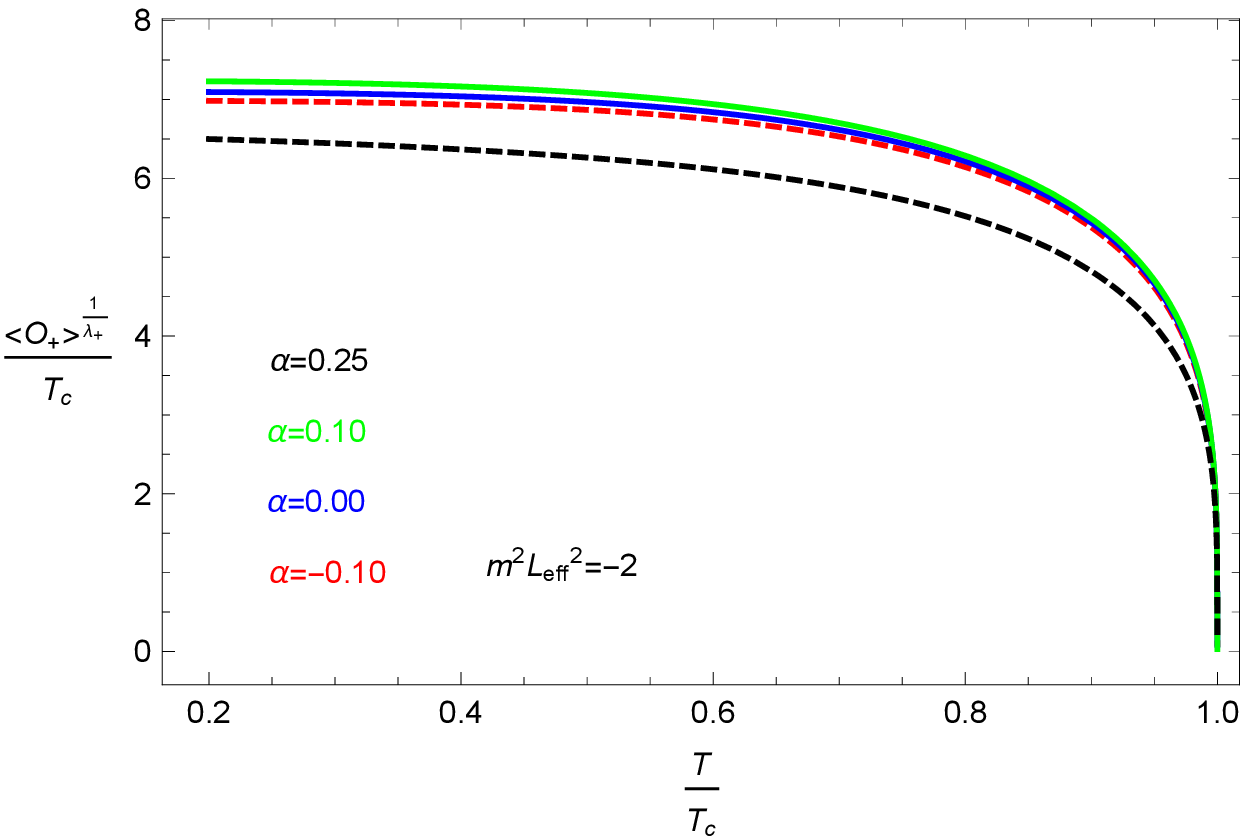}\\ \vspace{0.0cm}
\caption{\label{CondSWaveM2} (color online) The condensates of the scalar operator $O_{+}$ as a function of temperature for the fixed masses of the scalar field $m^2L^2=-2$ (left) and $m^{2}L_{\rm eff}^{2}=-2$ (right) with different Gauss-Bonnet parameters, i.e., $\alpha=-0.1$ (red and dashed), $0.0$ (blue), $0.1$ (green) and $0.25$ (black and dashed), respectively. }
\end{figure}

In Fig. \ref{CondSWaveM2}, we present the condensates of the scalar operator $O_{+}$ as a function of temperature with various Gauss-Bonnet parameters $\alpha$ for the fixed masses of the scalar field $m^2L^2=-2$ and $m^{2}L_{\rm eff}^{2}=-2$. It is found that, the curves have similar behavior to the BCS theory for different $\alpha$, where the condensate goes to a constant at zero temperature, which indicates that the s-wave holographic superconductors still exist even we consider Gauss-Bonnet correction terms to the standard ($2+1$)-dimensional holographic superconductor model built in \cite{HartnollPRL101}. Fitting these curves for small condensate, we observe that there is a square root behavior $\langle O_{+}\rangle\sim(1-T/T_{c})^{1/2}$, which is typical of second order phase transitions with the mean field critical exponent $1/2$ for all values of $\alpha$. Furthermore, the left panel of Fig. \ref{CondSWaveM2} shows that the higher correction term $\alpha$ makes the condensation gap larger for the operator $O_{+}$, which implies the higher Gauss-Bonnet correction will make the scalar hair more difficult to be developed. However, fixing the scalar field mass by $m^{2}L_{\rm eff}^{2}=-2$, we can see clearly from the right panel of Fig. \ref{CondSWaveM2} that there is an abnormal behavior in the Chern-Simons limit $\alpha=0.25$, which is completely different from that of the ($3+1$)-dimensional Gauss-Bonnet superconductors \cite{Gregory,Pan-Wang}. This means that the different choices of the mass of the scalar field will modify the effect of $\alpha$ on the behavior of the condensates even for the operator $O_{+}$.

\begin{figure}[ht]
\includegraphics[scale=0.65]{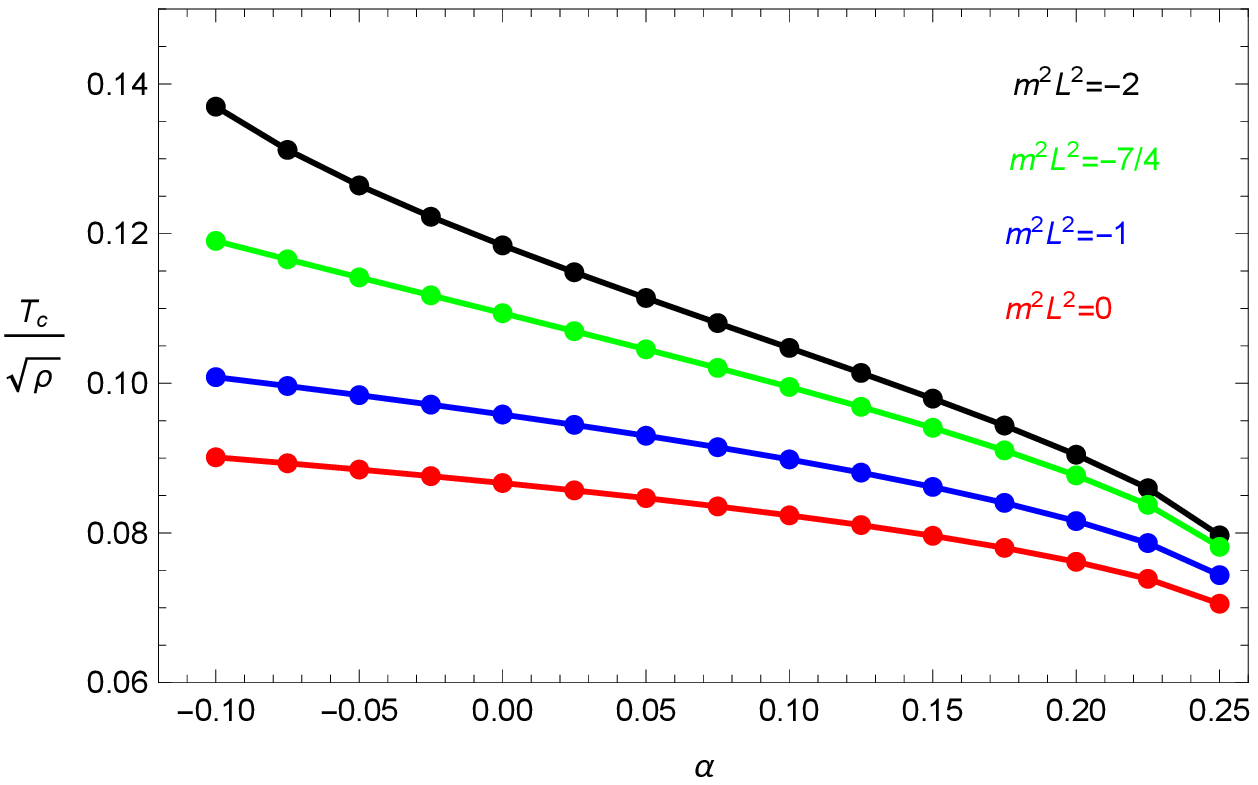}\vspace{0.0cm}
\includegraphics[scale=0.65]{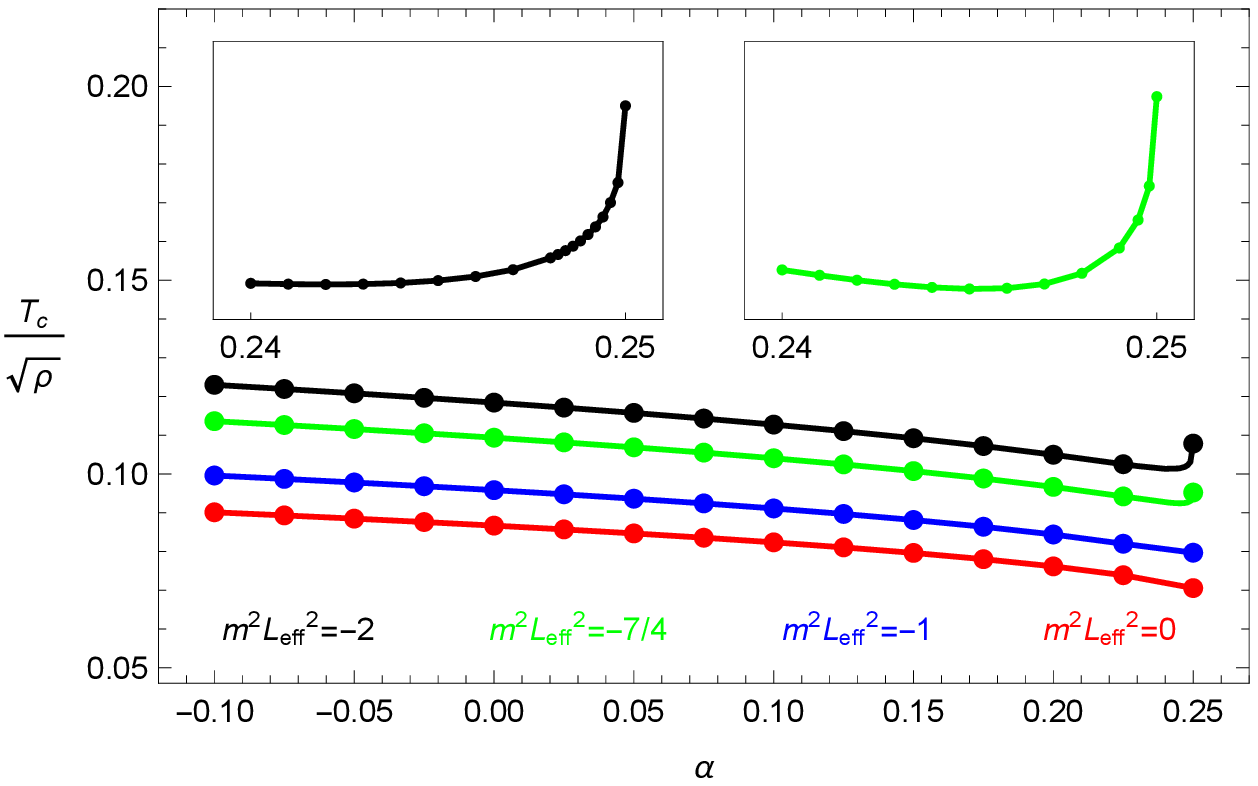}\\ \vspace{0.0cm}
\caption{\label{TcSWave} (color online) The critical temperature $T_{c}$ of the scalar operator $O_{+}$ as a function of the Gauss-Bonnet parameter for the fixed masses of the scalar field by choosing values of $m^2L^2$ (left) and $m^{2}L_{\rm eff}^{2}$ (right). }
\end{figure}

In order to further understand the findings of Fig. \ref{CondSWaveM2}, we exhibit the critical temperature $T_{c}$ as a function of the Gauss-Bonnet parameter $\alpha$ for the different choices of the mass of the scalar field in Fig. \ref{TcSWave}. Fixing the mass of the scalar field by choosing values of $m^{2}L^{2}$, from the left panel of Fig. \ref{TcSWave} we observe that the critical temperature $T_{c}$ decreases as $\alpha$ increases for different masses of the scalar field, which agrees well with the finding in the left panel of Fig. \ref{CondSWaveM2} and indicates that the higher curvature correction makes the condensate of the scalar operator $O_{+}$ harder. However, the story is different if we fix the mass of the scalar field by choosing values of $m^{2}L_{\rm eff}^{2}$. From the right panel of Fig. \ref{TcSWave}, we find that the effect of the Gauss-Bonnet parameter is more subtle: the critical temperature $T_{c}$ decreases with the increase of $\alpha$ for the large mass scale such as $m^{2}L_{\rm eff}^{2}=0$ and $-1$, but decreases first and then increases when $\alpha$ increases for the small mass scale such as $m^{2}L_{\rm eff}^{2}=-7/4$ and $-2$, which is consistent with the behavior of the scalar condensates for $m^{2}L_{\rm eff}^{2}=-2$ with various Gauss-Bonnet parameters in Fig. \ref{CondSWaveM2}. This behavior is reminiscent of that seen for the ($3+1$)-dimensional Gauss-Bonnet superconductors with backreactions, where the critical temperature first decreases then increases as the Gauss-Bonnet term tends towards the Chern-Simons value in a backreaction dependent fashion \cite{BarclayGregory}. Obviously, although how the curvature correction works in the holographic s-wave superconductors is still an open question, the ($2+1$)-dimensional Gauss-Bonnet superconductors exhibit a very interesting and different feature when compared to the higher-dimensional cases in the probe limit.

\subsubsection{Scalar operator $O_{-}$}

Now we are in a position to impose the condition $\psi_{+}=0$ and discuss the condensates of the scalar operator $O_{-}$. For concreteness, we set $m^2L^2=-2$ and $m^{2}L_{\rm eff}^{2}=-2$ in our calculation, where the choice of the Gauss-Bonnet parameter should satisfy the range $-9/4<m^{2}L_{\rm eff}^2<-5/4$ where both modes of the asymptotic values of the scalar fields are normalizable \cite{HorowitzPRD78}.

\begin{figure}[ht]
\includegraphics[scale=0.65]{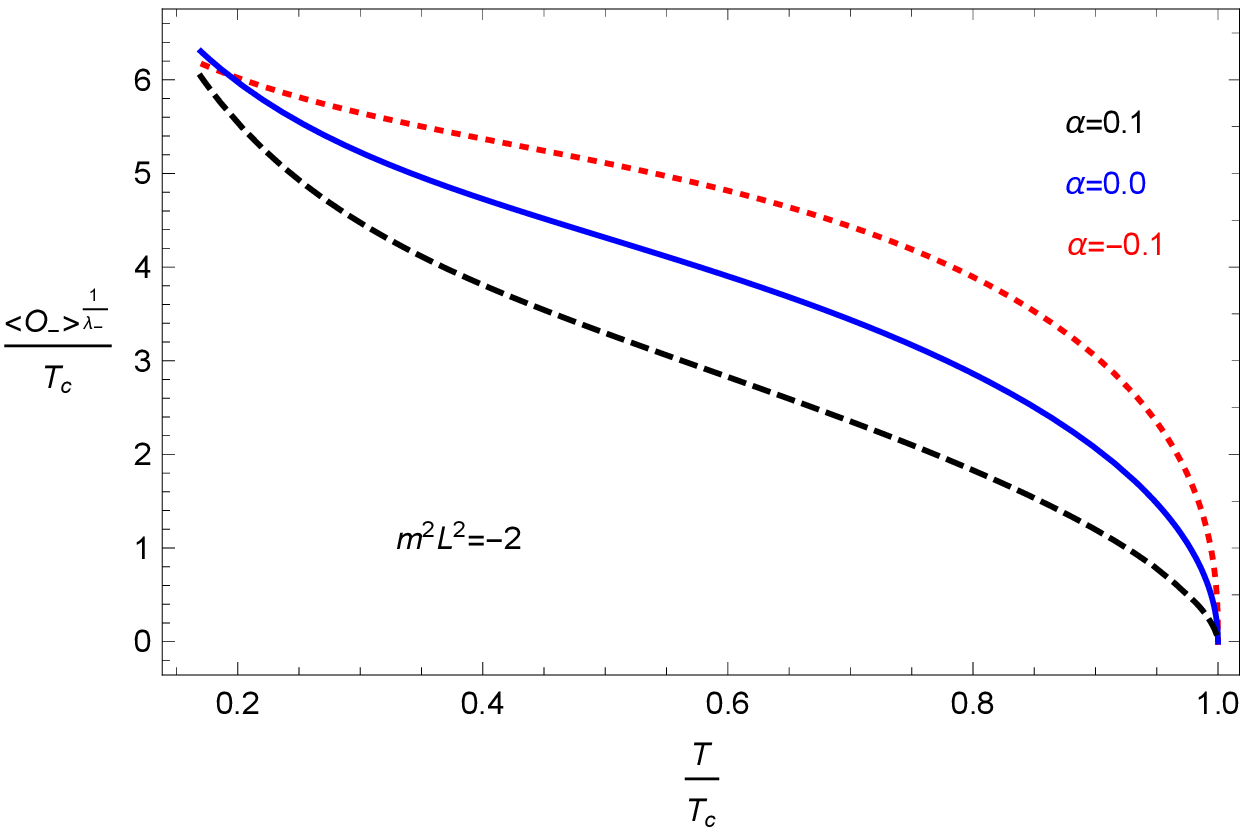}\vspace{0.0cm}
\includegraphics[scale=0.65]{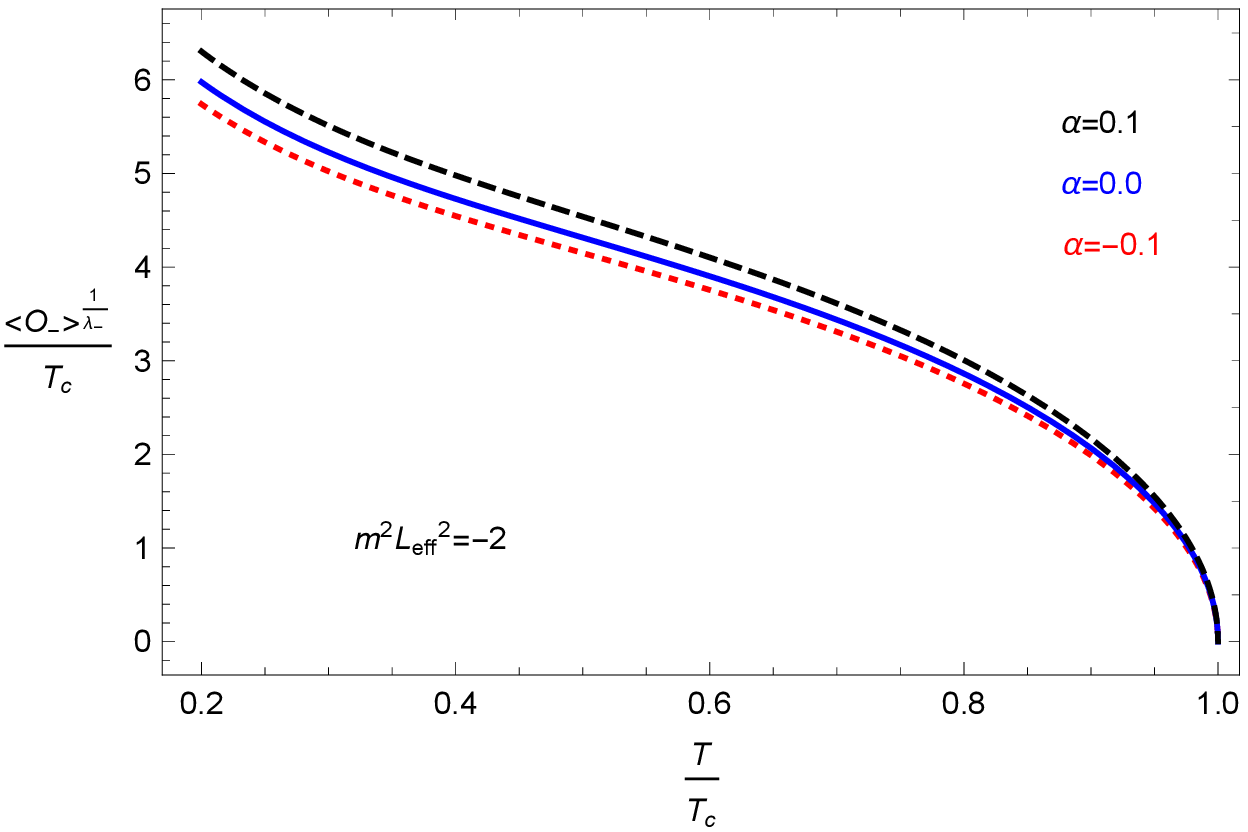}\\ \vspace{0.0cm}
\caption{\label{CondSWaveFuM2} (color online) The condensates of the scalar operator $O_{-}$ as a function of temperature for the fixed masses of the scalar field $m^2L^2=-2$ (left) and $m^{2}L_{\rm eff}^{2}=-2$ (right) with different Gauss-Bonnet parameters, i.e., $\alpha=-0.1$ (red and dashed), $0.0$ (blue) and $0.1$ (black and dashed), respectively. }
\end{figure}

\begin{figure}[ht]
\includegraphics[scale=0.75]{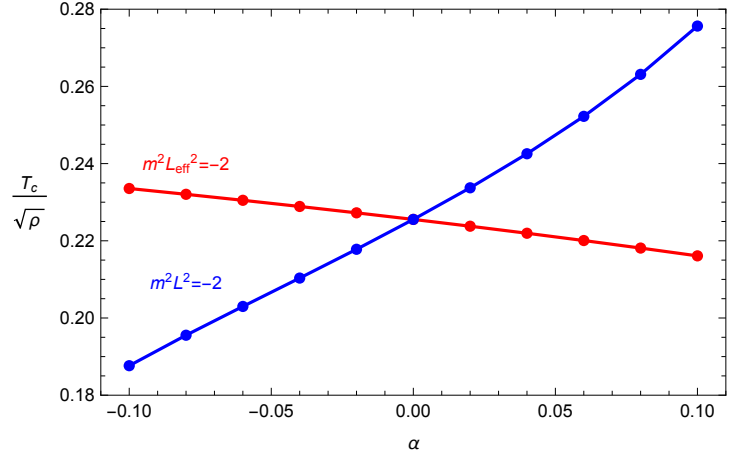}\\ \vspace{0.0cm}
\caption{\label{TcSWaveFu} (color online) The critical temperature $T_{c}$ of the scalar operator $O_{-}$ as a function of the Gauss-Bonnet parameter for the fixed masses of the scalar field $m^2L^2=-2$ (blue) and $m^{2}L_{\rm eff}^{2}=-2$ (red). }
\end{figure}

We give in Fig. \ref{CondSWaveFuM2} the condensates of the scalar operator $O_{-}$ as a function of temperature with various Gauss-Bonnet parameters $\alpha$ for the fixed masses $m^2L^2=-2$ and $m^{2}L_{\rm eff}^{2}=-2$, which will diverge at low temperatures, similar to that for the standard holographic superconductor model in the probe limit neglecting backreaction of the spacetime \cite{HartnollPRL101}. More importantly, from this figure, we find that the condensate of the operator $O_{-}$ for different choices of the mass of the scalar field has completely different behavior as the Gauss-Bonnet parameter is changing, which is in agreement with the result obtained in the Fig. \ref{TcSWaveFu}, where the Gauss-Bonnet term has completely different effects on the critical temperature $T_{c}$ for $m^2L^2=-2$ and $m^{2}L_{\rm eff}^{2}=-2$. Considering the correct consistent influence due to the Gauss-Bonnet parameter in various condensates for all dimensions \cite{Pan-Wang,Pan-Jing-Wang-Soliton} and the pure effect of the curvature correction on the critical temperature $T_{c}$ for the fixed mass $m^{2}L^{2}=m^{2}L_{\rm eff}^{2}=0$ in Fig. \ref{TcSWave}, i.e., the increase of $\alpha$ results in the decrease of $T_{c}$, we argue that it is more appropriate to choose the mass of the scalar field by selecting the values of $m^{2}L_{\rm eff}^{2}$ for the Gauss-Bonnet superconductors even in ($2+1$)-dimensions. Therefore, the higher curvature corrections make it harder for the condensate of the scalar operator $O_{-}$ to form.

\subsection{Conductivity}

In order to calculate the conductivity, we consider the perturbed Maxwell field $\delta
A_{x}=A_{x}(r)e^{-i\omega t}dx$, which results in the equation of motion
\begin{eqnarray}
A_{x}^{\prime\prime}+\frac{f^\prime}{f}A_{x}^\prime
+\left(\frac{\omega^2}{f^2}-\frac{2q^{2}\psi^{2}}{f}\right)A_{x}=0. \label{ConductivityEquation}
\end{eqnarray}
With the ingoing wave boundary condition near the horizon $A_{x}(r)\sim f(r)^{-i\omega/(3r_{+})}$, and the behavior in the asymptotic AdS region
\begin{eqnarray}
A_{x}=A^{(0)}_{x}+\frac{A^{(1)}_{x}}{r},
\end{eqnarray}
from Ohm's law we can obtain the conductivity of the dual superconductor \cite{HartnollPRL101}
\begin{eqnarray}\label{SWConductivity}
\sigma=-\frac{iA^{(1)}_{x}}{\omega A^{(0)}_{x}}.
\end{eqnarray}
We will focus on the case of the scalar operator $O_{+}$ for the fixed scalar field masses $m^{2}L_{\rm eff}^{2}=-2$ and $0$, and study the effect of the Gauss-Bonnet correction term on the conductivity by solving Eq. (\ref{ConductivityEquation}) numerically.

\begin{figure}[H]
\includegraphics[scale=0.425]{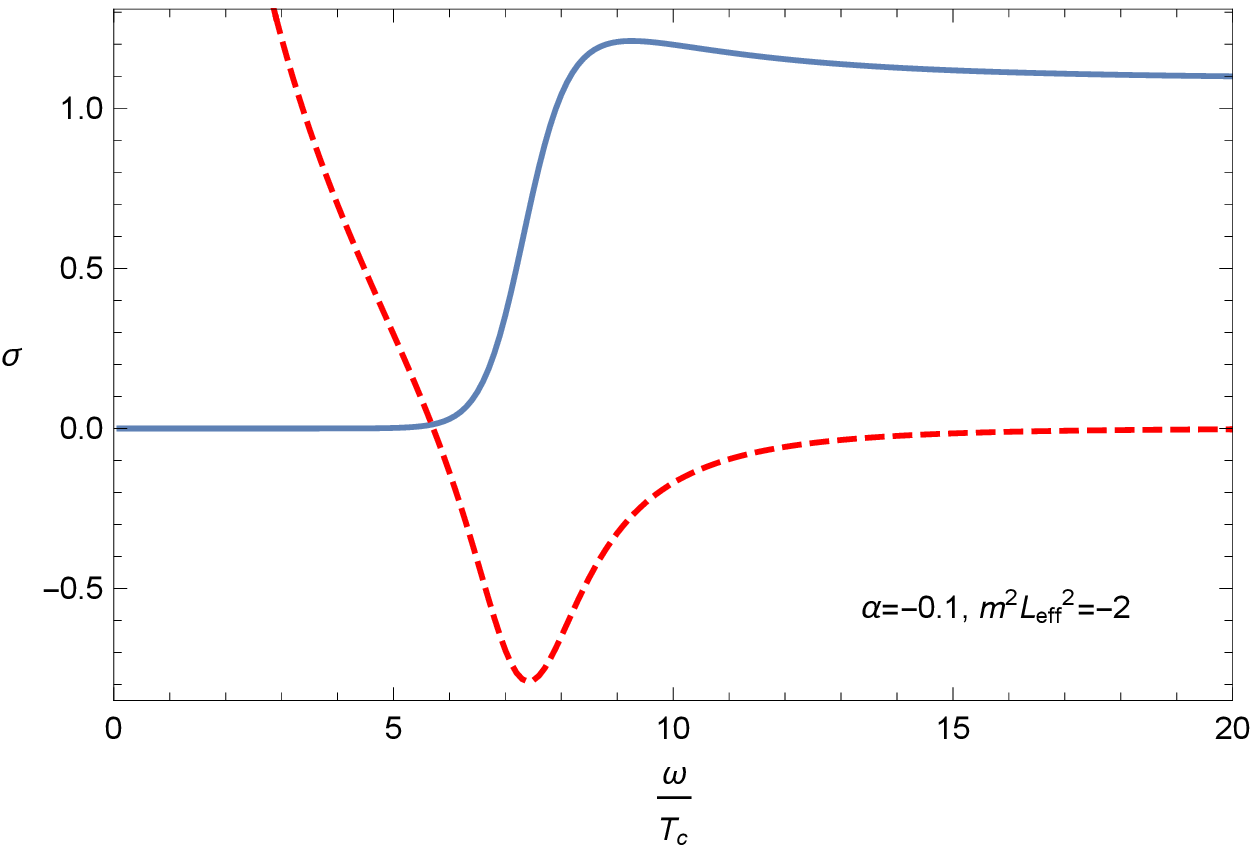}\hspace{0.2cm}%
\includegraphics[scale=0.425]{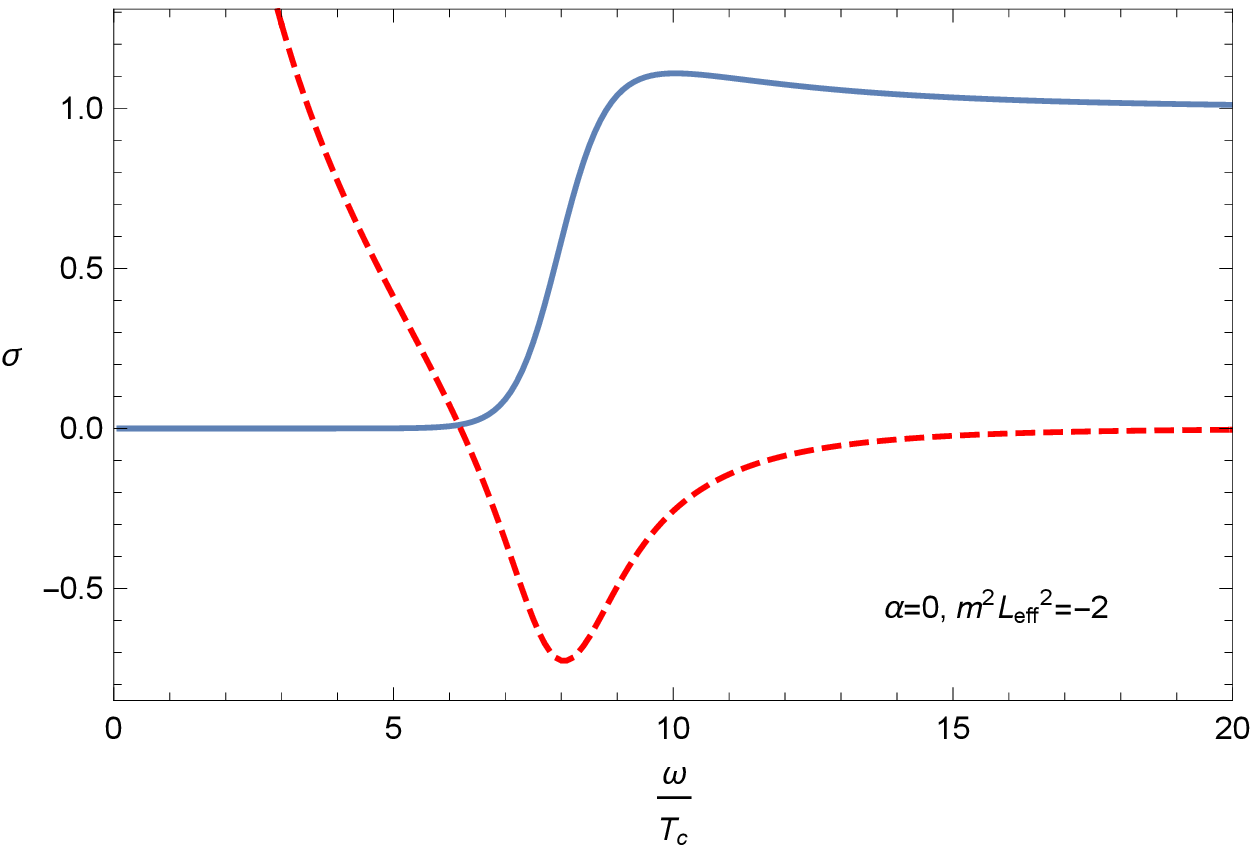}\hspace{0.2cm}%
\includegraphics[scale=0.425]{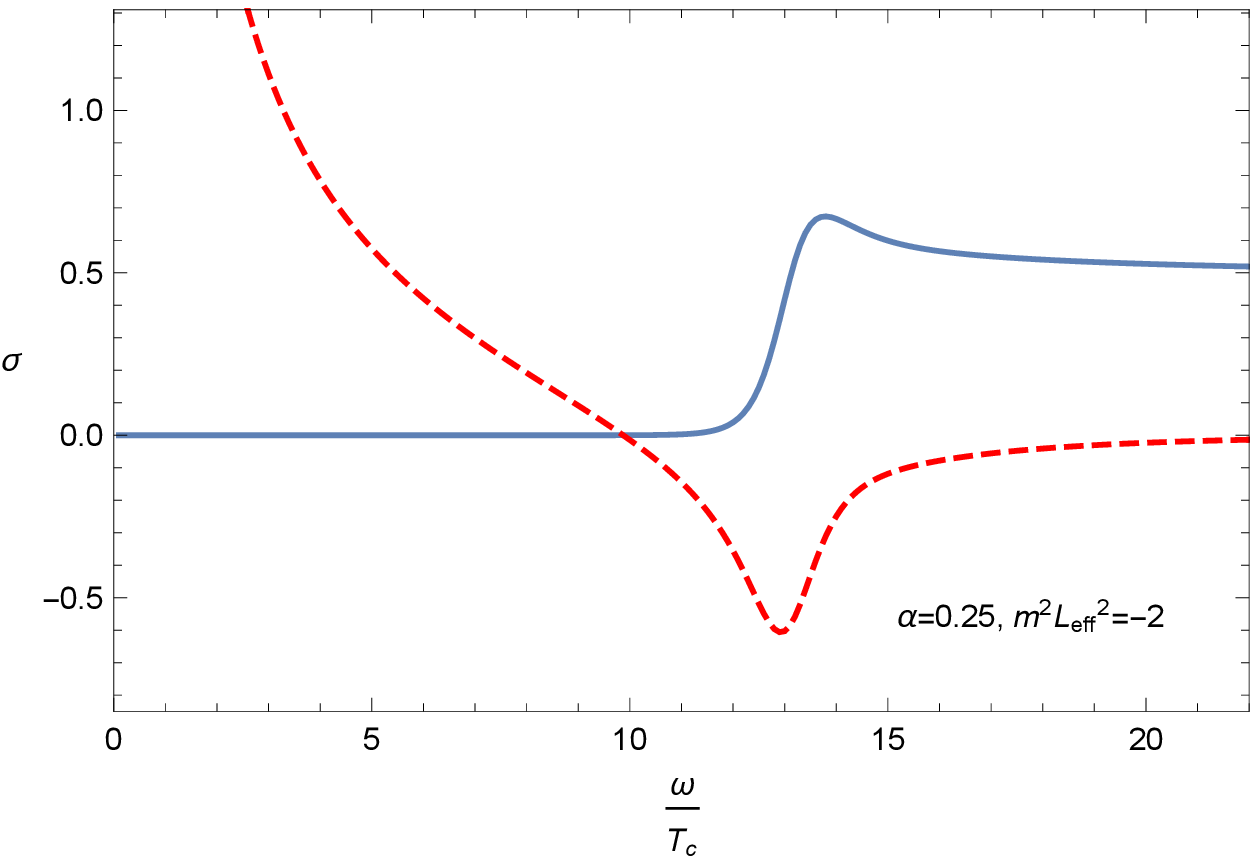}\\ \vspace{0.0cm}
\includegraphics[scale=0.425]{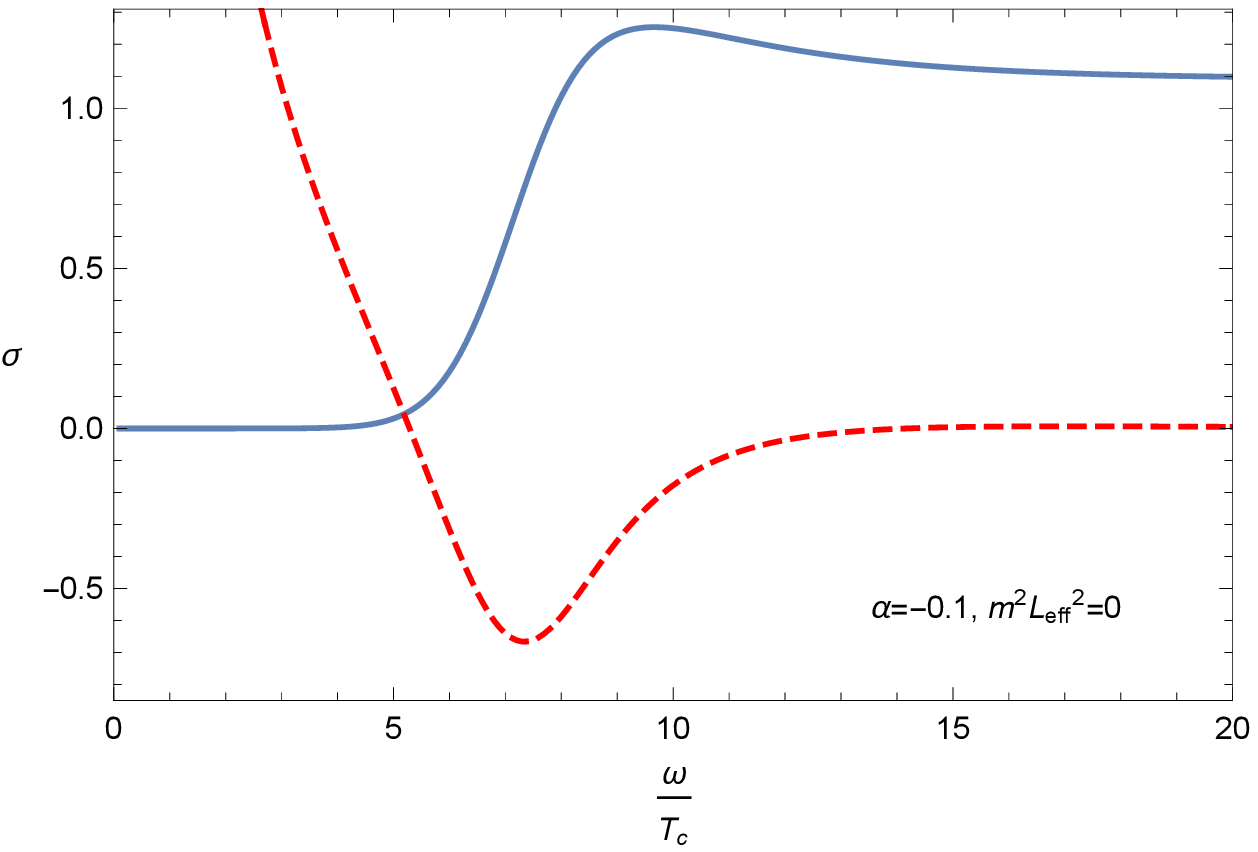}\hspace{0.2cm}%
\includegraphics[scale=0.425]{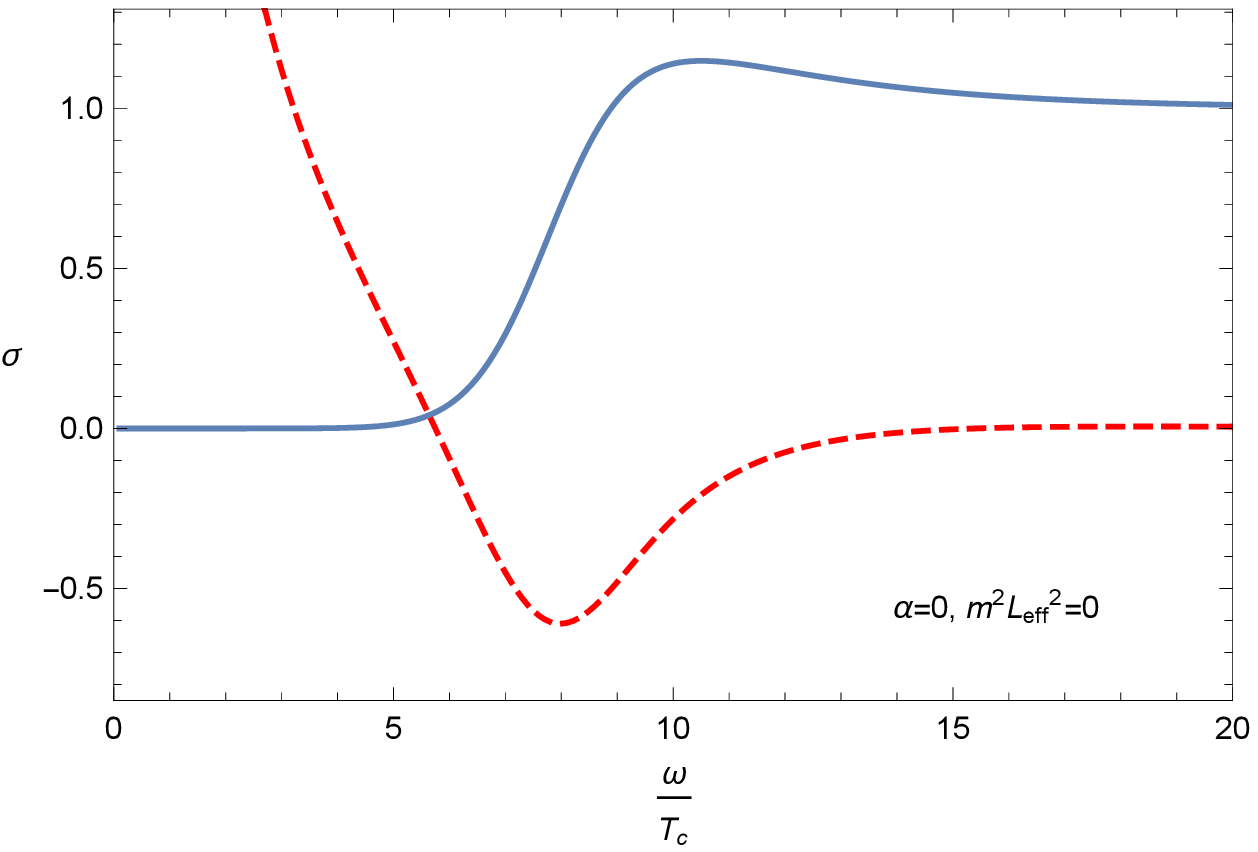}\hspace{0.2cm}%
\includegraphics[scale=0.425]{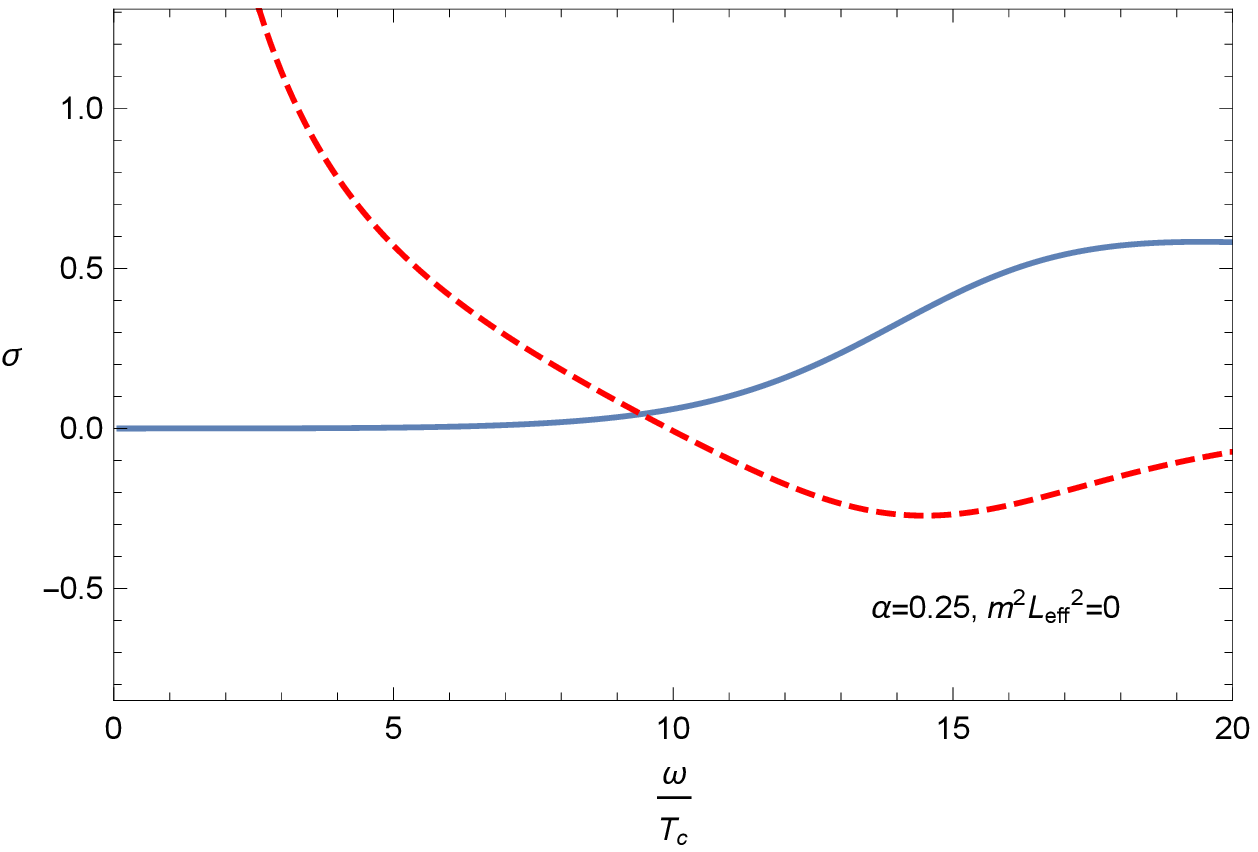}\\ \vspace{0.0cm}
\caption{\label{SWaveConductivity} (color online) Conductivity of the ($2+1$)-dimensional s-wave superconductors for different values of the Gauss-Bonnet parameter with the fixed scalar field masses $m^{2}L_{\rm eff}^{2}=-2$ and $0$. In each panel, the blue (solid) line and red (dashed) line represent the real part and imaginary part of the conductivity $\sigma(\omega)$ respectively.}
\end{figure}

In Fig. \ref{SWaveConductivity} we present the frequency dependent conductivity for $\alpha=-0.1$, $0$ and $0.25$ with the fixed masses $m^{2}L_{\rm eff}^{2}=-2$ and $0$ at temperatures $T/T_{c}\approx0.2$. We can see clearly a gap in the conductivity with the gap frequency $\omega_{g}$ which can be defined as the frequency minimizing $Im[\sigma(\omega)]$ for all cases
considered here \cite{HorowitzPRD78}. For the same mass of the scalar field, we observe that with the increase of the Gauss-Bonnet parameter $\alpha$, the gap frequency $\omega_{g}$ becomes larger, just as in the ($3+1$)-dimensional Gauss-Bonnet superconductors \cite{Gregory,Pan-Wang}. Also, for increasing absolute value of $\alpha$, we have larger deviations from the value $\omega_g/T_c\approx 8$, especially in the case of the Chern-Simons limit $\alpha=0.25$ where $\omega_g/T_c\approx 13.0$ for $m^{2}L_{\rm eff}^{2}=-2$ and $\omega_g/T_c\approx 14.5$ for $m^{2}L_{\rm eff}^{2}=0$. This shows that the higher curvature corrections really change the universal relation in the gap frequency $\omega_g/T_c\approx 8$ \cite{HorowitzPRD78} for ($2+1$)-dimensional superconductors.

\section{p-Wave holographic superconductors}

In the previous section, we have investigated the influence of the curvature correction on the ($2+1$)-dimensional s-wave superconductors, which shows that the Gauss-Bonnet terms have more subtle effects on the condensates of the scalar field and change the ratio in the gap frequency $\omega_g/T_c\approx 8$. In this section, we will construct the p-wave holographic superconductors in the 4D Gauss-Bonnet-AdS black hole and analyze the influence of the curvature correction on these models.

\subsection{Condensates of the vector field}

We start with the action of the Einstein-Maxwell-complex vector model \cite{CaiPWave-1,CaiPWave-2}
\begin{eqnarray}\label{pWaveSystem}
S=\int
d^{4}x\sqrt{-g}\left[-\frac{1}{4}F_{\mu\nu}F^{\mu\nu}-\frac{1}{2}
(D_\mu\rho_\nu-D_\nu\rho_\mu)^{\dag}(D^{\mu}\rho^{\nu}-D^{\nu}\rho^{\mu})
-m^2\rho_{\mu}^{\dag}\rho^{\mu}+iq\gamma\rho_{\mu}\rho_{\nu}^{\dag}F^{\mu\nu}\right],
\end{eqnarray}
where $m$ and $q$ represent the mass and charge of the vector field $\rho_\mu$, and $D_\mu=\nabla_\mu-iqA_\mu$ is the covariant derivative. The last term, which describes the interaction between the vector field $\rho_\mu$ and gauge field $A_\mu$, will not play any role since we will only consider the case without an external magnetic field. Taking the ansatz for the matter fields $\rho_{\nu}dx^{\nu}=\rho_{x}(r)dx$ and $A_{\nu}dx^{\nu}=A_{t}(r)dt$ \cite{CaiPWave-1,CaiPWave-2}, we obtain the following equations of motion
\begin{eqnarray}
\rho_{x}^{\prime\prime}+\frac{f^\prime}{f}\rho_{x}^{\prime}
+\left(\frac{q^{2}A_{t}^2}{f^2}-\frac{m^2}{f}\right)\rho_{x}=0,
\label{rhoxPWave}
\end{eqnarray}
\begin{eqnarray}
A_t^{\prime\prime}+\frac{2}{r}A_t^{\prime}-\frac{2q^2\rho_{x}^2}{r^2f}A_t=0, \label{AtPWave}
\end{eqnarray}
where the prime denotes the derivative with respect to $r$.

We impose the appropriate boundary conditions for $\rho_{x}(r)$ and $A_{t}(r)$ to get the solutions in the superconducting phase. At the horizon, we can easily get the boundary conditions $\rho_{x}(r_{+})=f^\prime(r_{+})\rho_{x}^\prime(r_{+})/m^{2}$ and $A_{t}(r_{+})=0$. At the asymptotic boundary, we have
\begin{eqnarray}
\rho_{x}=\frac{\rho_{x-}}{r^{\lambda_{-}}}+\frac{\rho_{x+}}{r^{\lambda_{+}}}\,,\hspace{0.5cm}
A_{t}=\mu-\frac{\rho}{r}\,, \label{infinityPWave}
\end{eqnarray}
with the characteristic exponents
\begin{eqnarray}
\lambda_\pm=\frac{1}{2}\left(1\pm\sqrt{1+4m^{2}L_{\rm eff}^2}\right), \label{PWcharacteristicexponent}
\end{eqnarray}
where $\rho_{x-}$ and $\rho_{x+}$ are interpreted as the source and the vacuum expectation value of the vector operator $O_{x}$ in the dual field theory according to the AdS/CFT correspondence, respectively. It should be noted that the mass has a lower bound $m^{2}_{BF}=-1/(4L_{\rm eff}^{2})$ with $\lambda_{+}=\lambda_{-}=1/2$. Since we require that the condensate appears spontaneously, we will impose the boundary condition $\rho_{x-}=0$ and use $\lambda$ to denote $\lambda_{+}$ in the following analysis.

Considering that Eqs. (\ref{rhoxPWave}) and (\ref{AtPWave}) are invariant with respect to the following scaling transformations
\begin{eqnarray}\label{PWavesymmetry}
&&r\rightarrow\beta r,~~~(t, x, y)\rightarrow \beta^{-1}(t, x, y),~~~q\rightarrow q,~~~(\rho_{x},A_{t})\rightarrow\beta(\rho_{x},A_{t}),\nonumber \\
&&(T,\mu)\rightarrow\beta(T,\mu),~~~\rho\rightarrow\beta^{2}\rho,~~~\rho_{x+}\rightarrow\beta^{1+\lambda_{+}}\rho_{x+},
\end{eqnarray}
with a positive number $\beta$. In what follows, we will use them to set $r_{+}=1$ and $q=1$, just as in the previous section.

\begin{figure}[ht]
\includegraphics[scale=0.65]{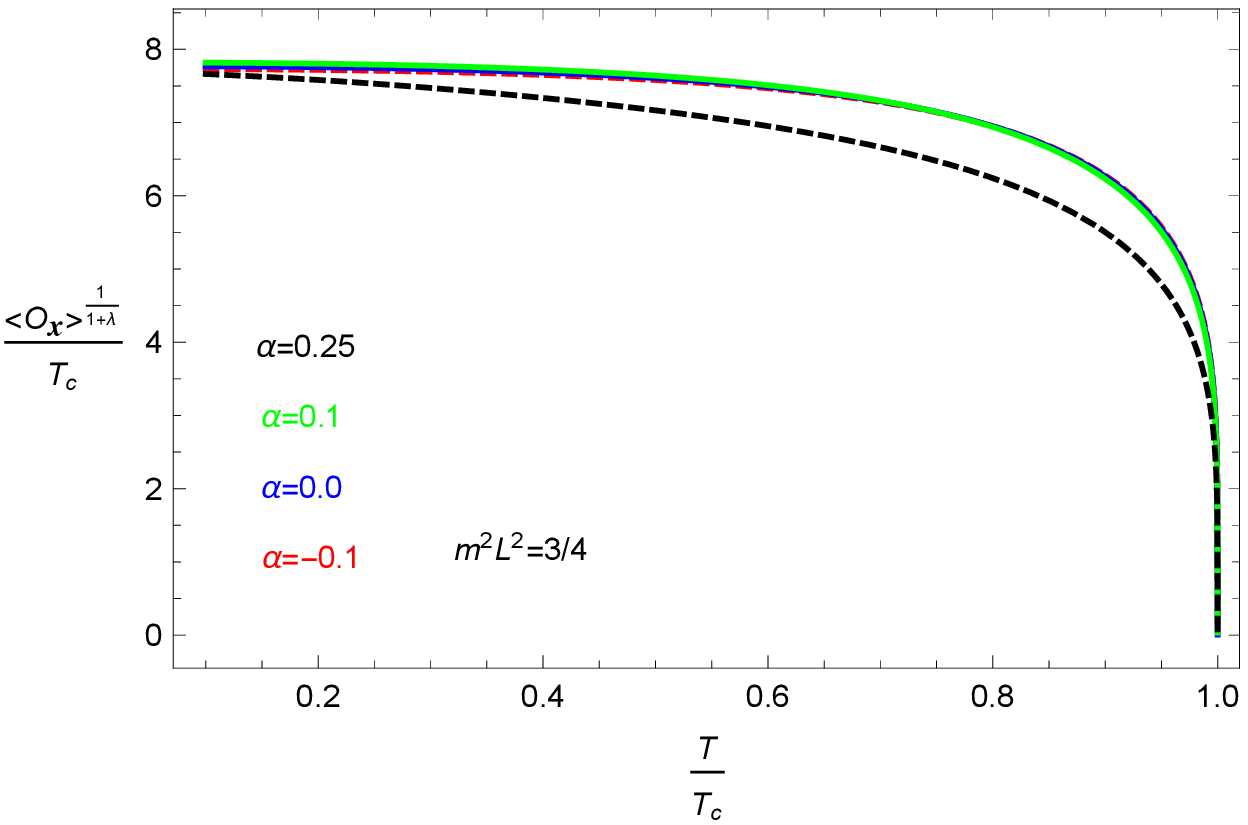}\vspace{0.0cm}
\includegraphics[scale=0.65]{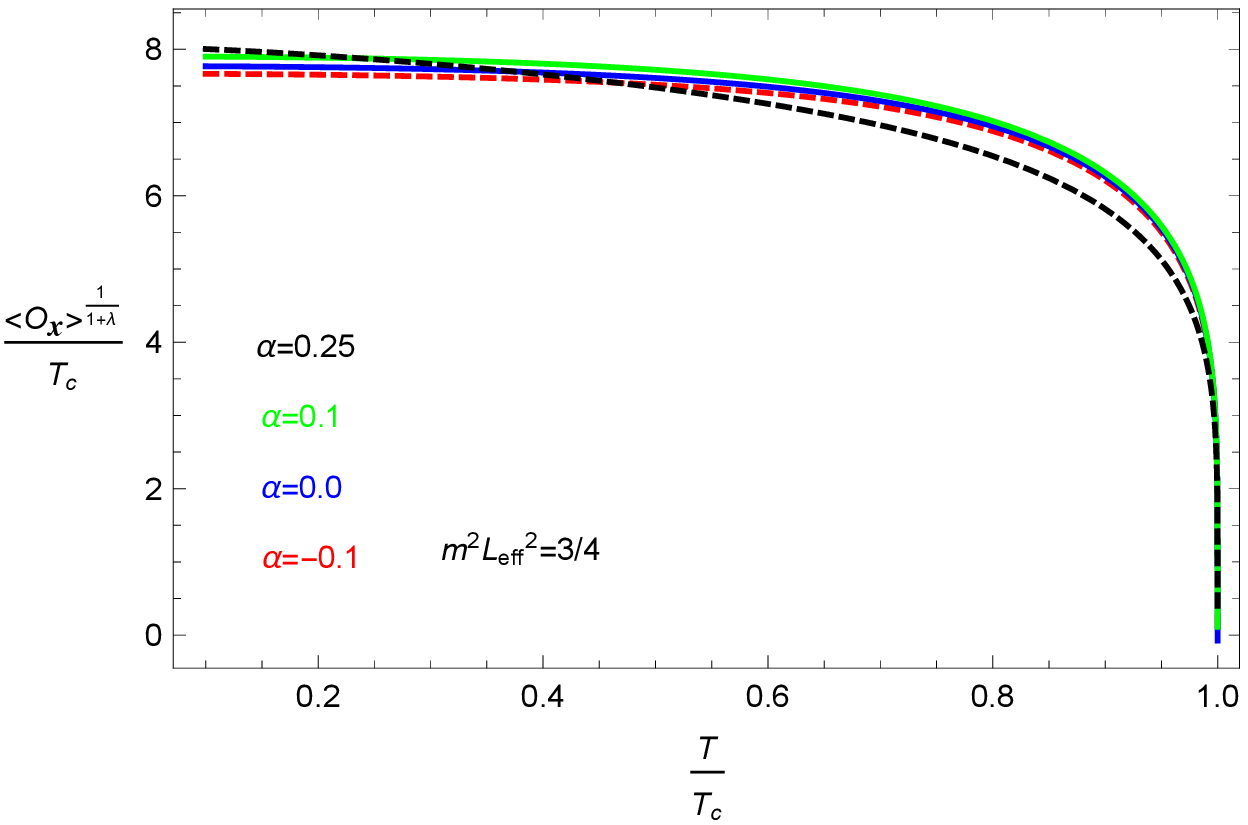}\\ \vspace{0.0cm}
\caption{\label{CondPWaveM2} (color online) The condensates of the vector operator $O_{x}$ as a function of temperature for the masses of the vector field $m^2L^2=3/4$ (left) and $m^{2}L_{\rm eff}^{2}=3/4$ (right). The four lines from bottom to top correspond to increasing Gauss-Bonnet parameter, i.e., $\alpha=-0.1$ (red and dashed), $0.0$ (blue), $0.1$ (green) and $0.25$ (black and dashed), respectively. }
\end{figure}

\begin{figure}[ht]
\includegraphics[scale=0.65]{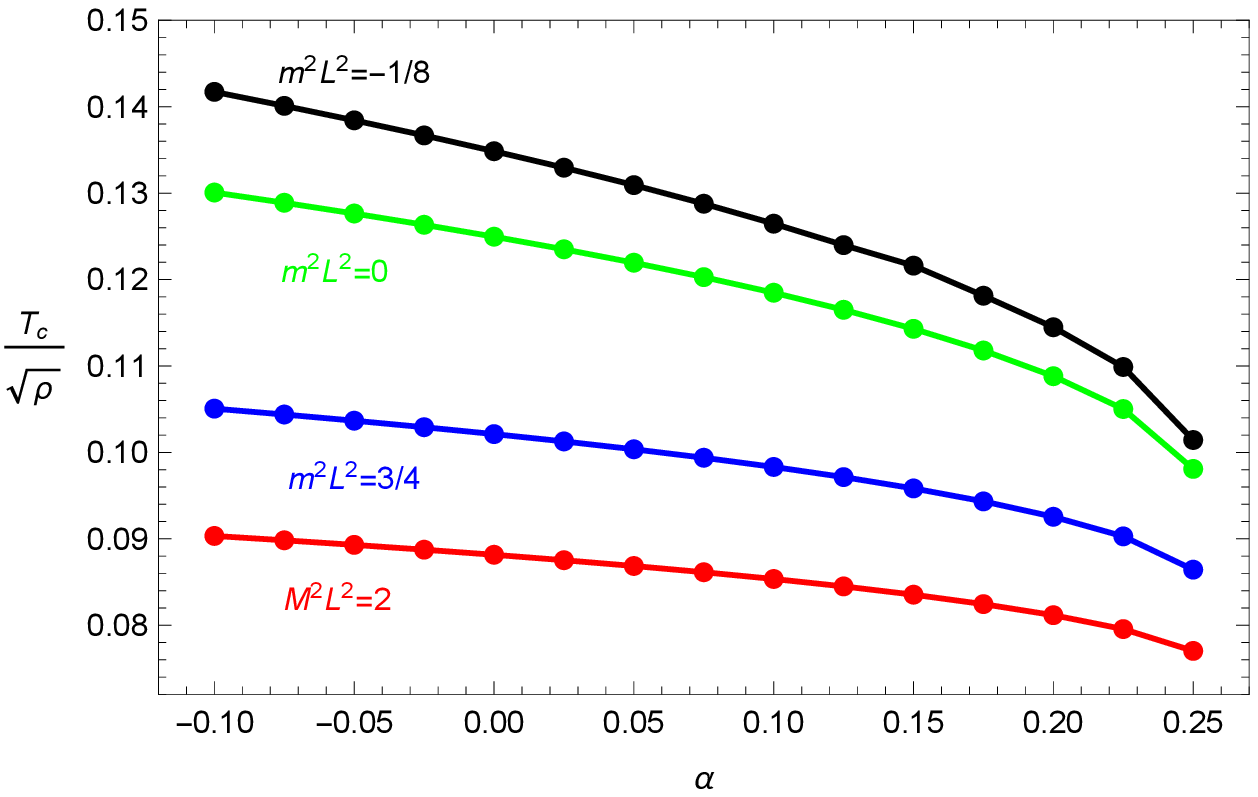}\vspace{0.0cm}
\includegraphics[scale=0.65]{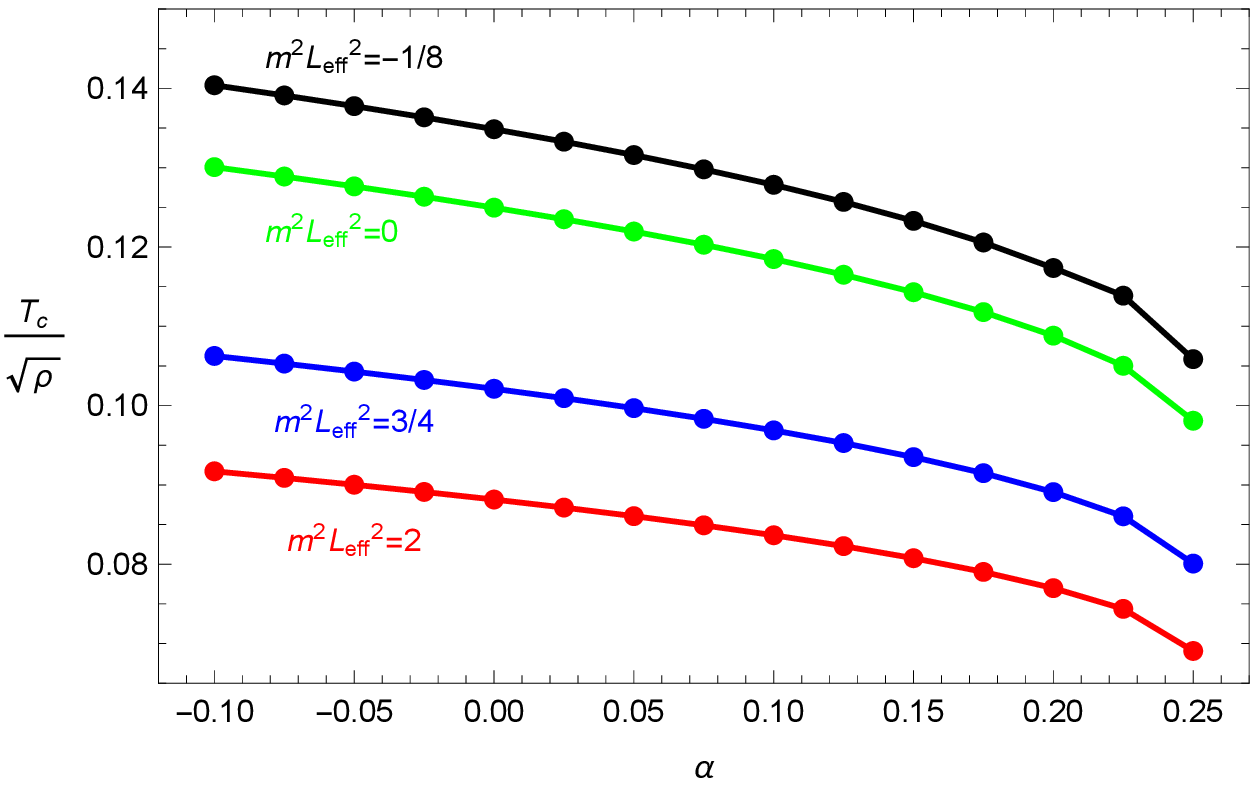}\\ \vspace{0.0cm}
\caption{\label{TcPWave} (color online) The critical temperature $T_{c}$ of the vector operator $O_{x}$ as a function of the Gauss-Bonnet parameter for the fixed masses of the vector field by choosing values of $m^2L^2$ (left) and $m^{2}L_{\rm eff}^{2}$ (right). }
\end{figure}

In Fig. \ref{CondPWaveM2}, we plot the condensates of the vector operator $O_{x}$ as a function of temperature with various Gauss-Bonnet corrections $\alpha$ for the fixed masses of the vector field $m^2L^2=3/4$ and $m^{2}L_{\rm eff}^{2}=3/4$. It is shown that, for all cases considered here, the behavior of the each curve is similar to that of the p-wave holographic
superconductor in the literature, which suggests that the 4D Gauss-Bonnet black-hole solution with a non-trivial vector field can describe a superconducting phase. Also, the transition is of the second order and the condensate
approaches zero as $\langle O_{x}\rangle\sim(1-T/T_{c})^{1/2}$ by fitting these curves for small condensate. Obviously, regardless of the choice of the mass of the vector field by selecting the value of $m^{2}L^{2}$ or $m^{2}L_{\rm eff}^{2}$, we see that the increase of the Gauss-Bonnet parameter $\alpha$ results in the increase of the condensation gap, which agrees well with the result given in the Fig. \ref{TcPWave}, where the critical temperature $T_{c}$ decreases as $\alpha$ increases for all vector field masses chosen by fixing $m^{2}L^{2}$ and $m^{2}L_{\rm eff}^{2}$. This means that the higher curvature corrections make it harder for the vector field to condense in the ($2+1$)-dimensional superconductors, which is quite different from the findings obtained in Fig. \ref{CondSWaveM2} for the s-wave superconductors where the different choices of the mass of the scalar field will modify the effect of the Gauss-Bonnet corrections on the behavior of the condensates.

\subsection{Conductivity}

Considering the perturbed Maxwell field $\delta A_{y}=A_{y}(r)e^{-i\omega t}dy$, we obtain the equation of motion
\begin{eqnarray}
A_{y}^{\prime\prime}+\frac{f^\prime}{f}A_{y}^\prime
+\left(\frac{\omega^2}{f^2}-\frac{2q^{2}\rho_{x}^{2}}{r^{2}f}\right)A_{y}=0, \label{PWaveConductivityEquation}
\end{eqnarray}
which has the ingoing wave boundary condition near the horizon $A_{y}(r)\sim f(r)^{-i\omega/(3r_{+})}$, and the general behavior near the asymptotic AdS boundary
\begin{eqnarray}
A_{y}=A^{(0)}_{y}+\frac{A^{(1)}_{y}}{r}.
\end{eqnarray}
Using the AdS/CFT dictionary, we get the conductivity \cite{CaiPWave-1}
\begin{eqnarray}\label{PWConductivity}
\sigma=-\frac{iA^{(1)}_{y}}{\omega A^{(0)}_{y}}.
\end{eqnarray}
For clarity, we choose the masses of the vector field $m^{2}L_{\rm eff}^{2}=3/4$ and $0$ in our calculation.

\begin{figure}[H]
\includegraphics[scale=0.425]{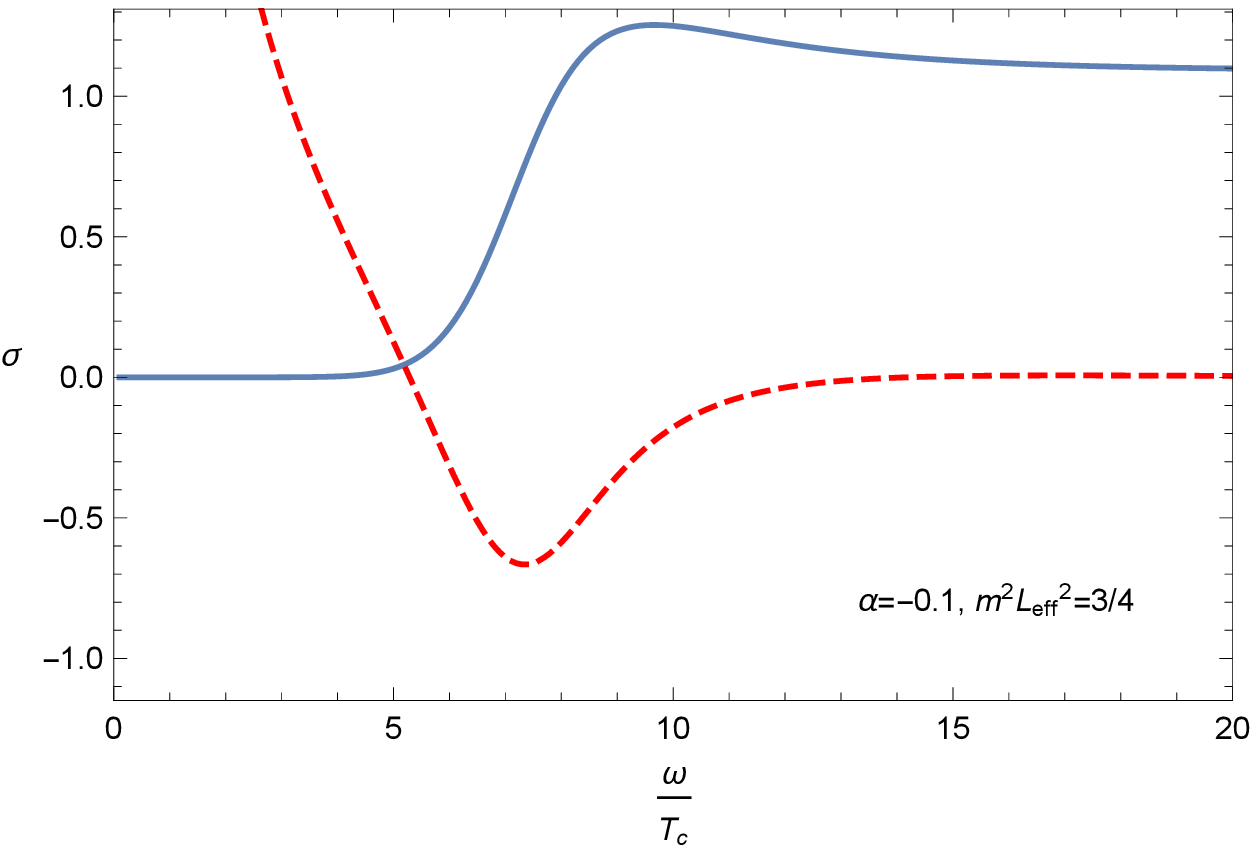}\hspace{0.2cm}%
\includegraphics[scale=0.425]{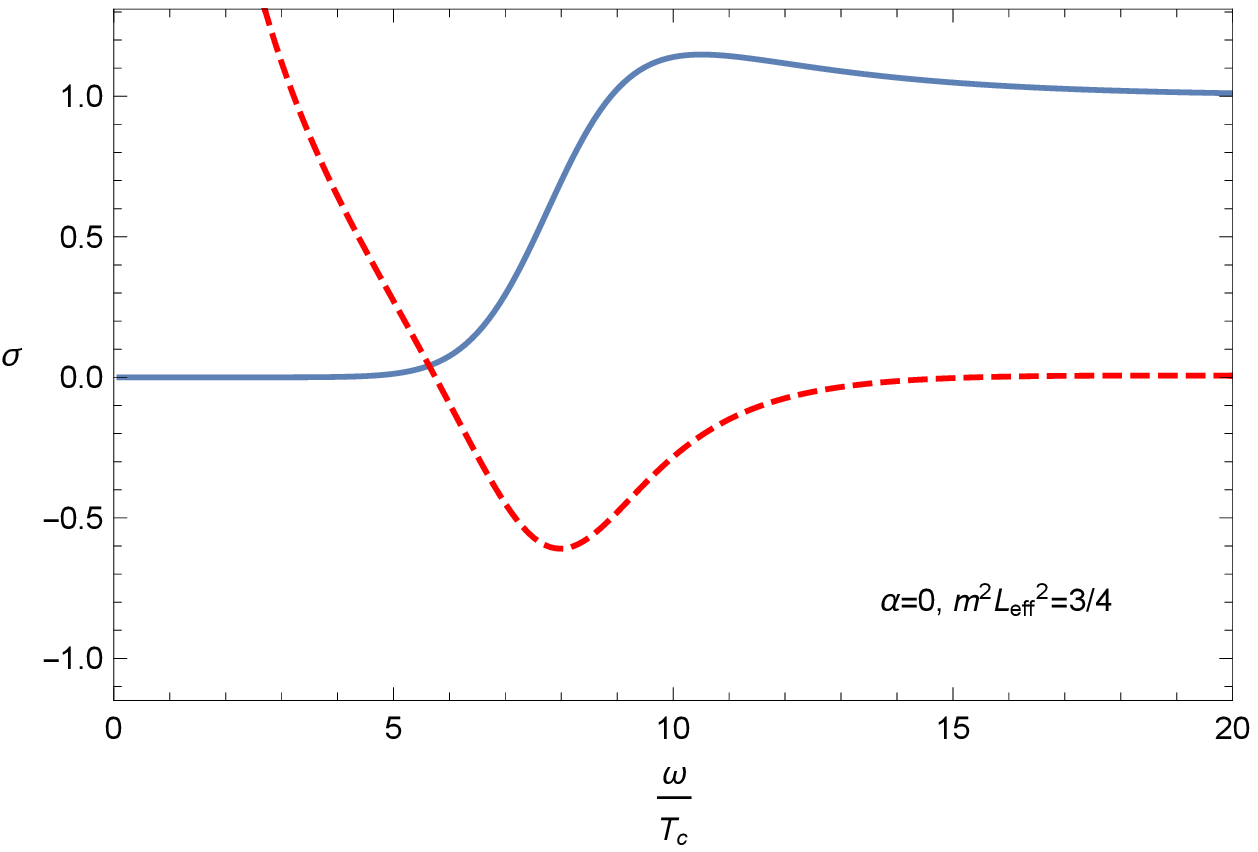}\hspace{0.2cm}%
\includegraphics[scale=0.425]{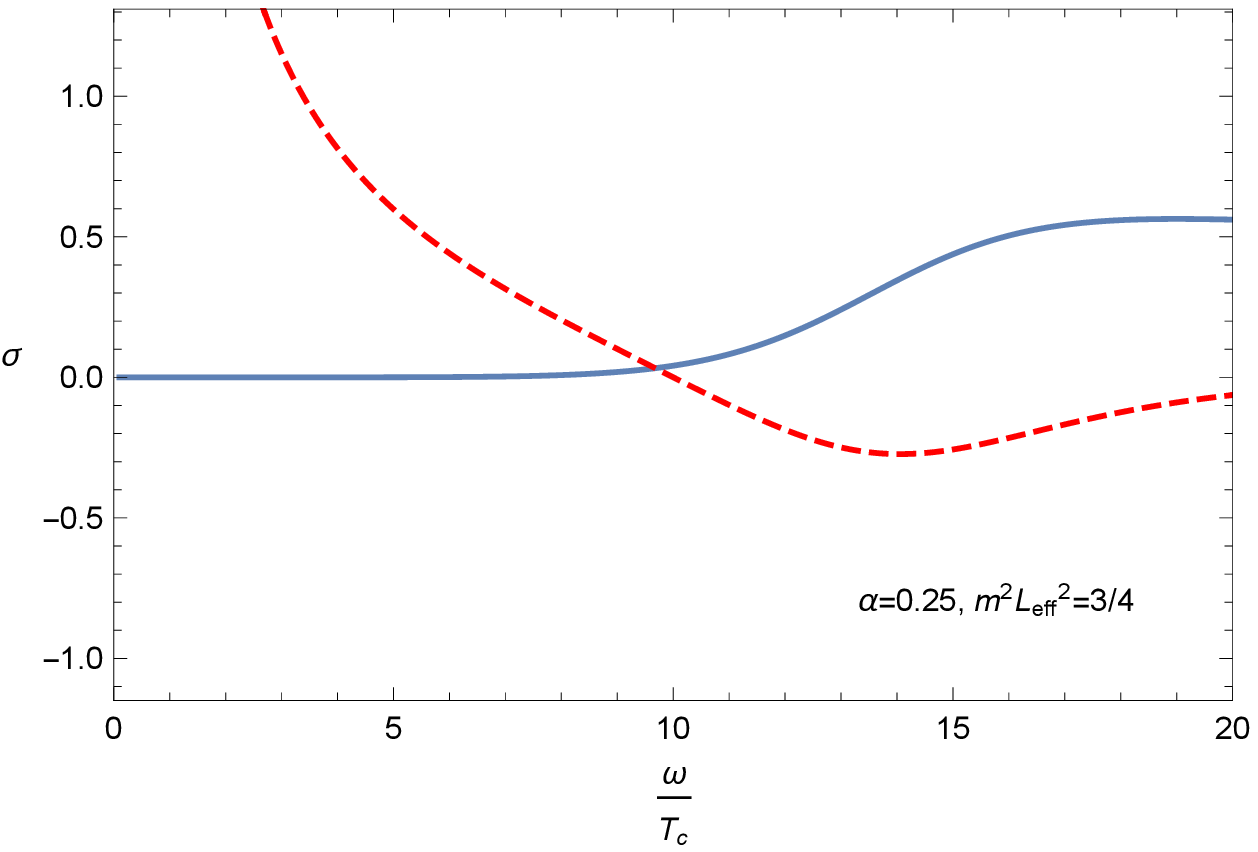}\\ \vspace{0.0cm}
\includegraphics[scale=0.425]{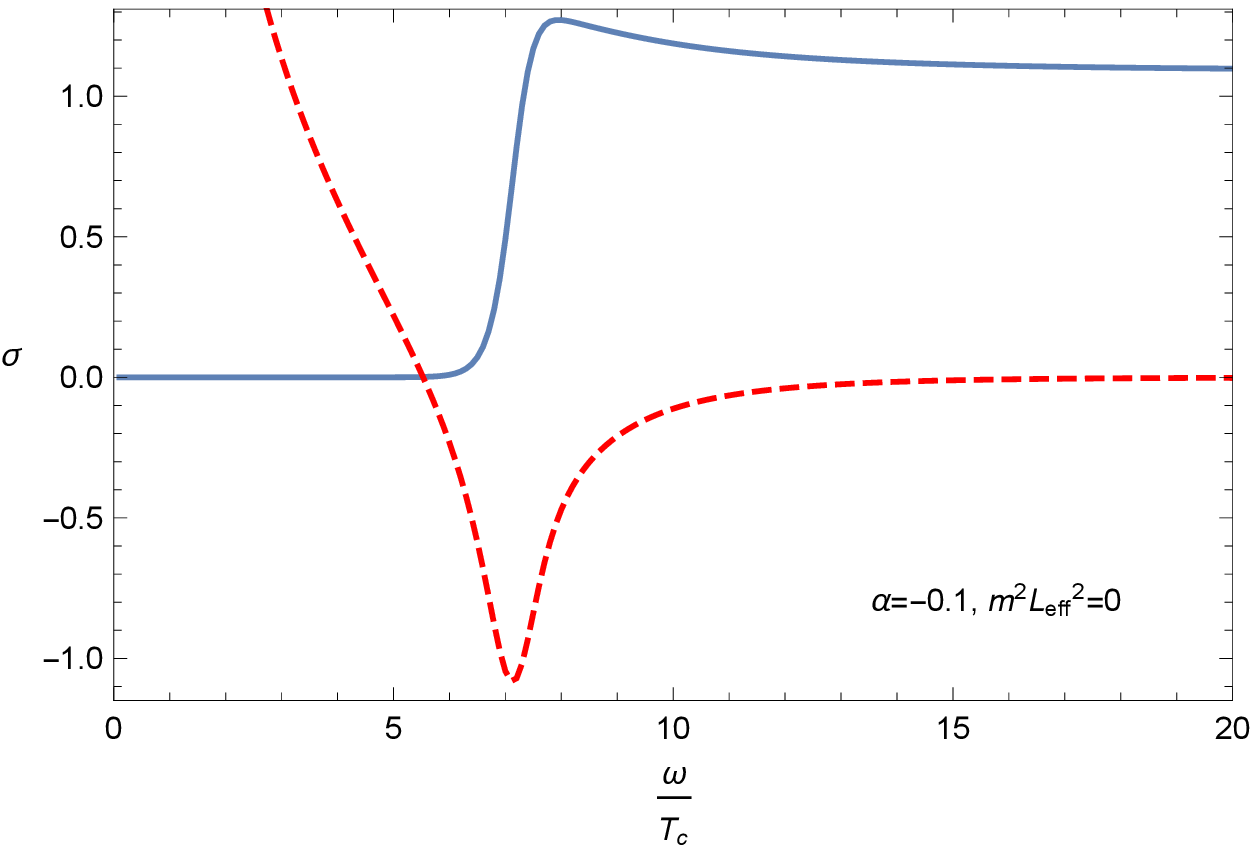}\hspace{0.2cm}%
\includegraphics[scale=0.425]{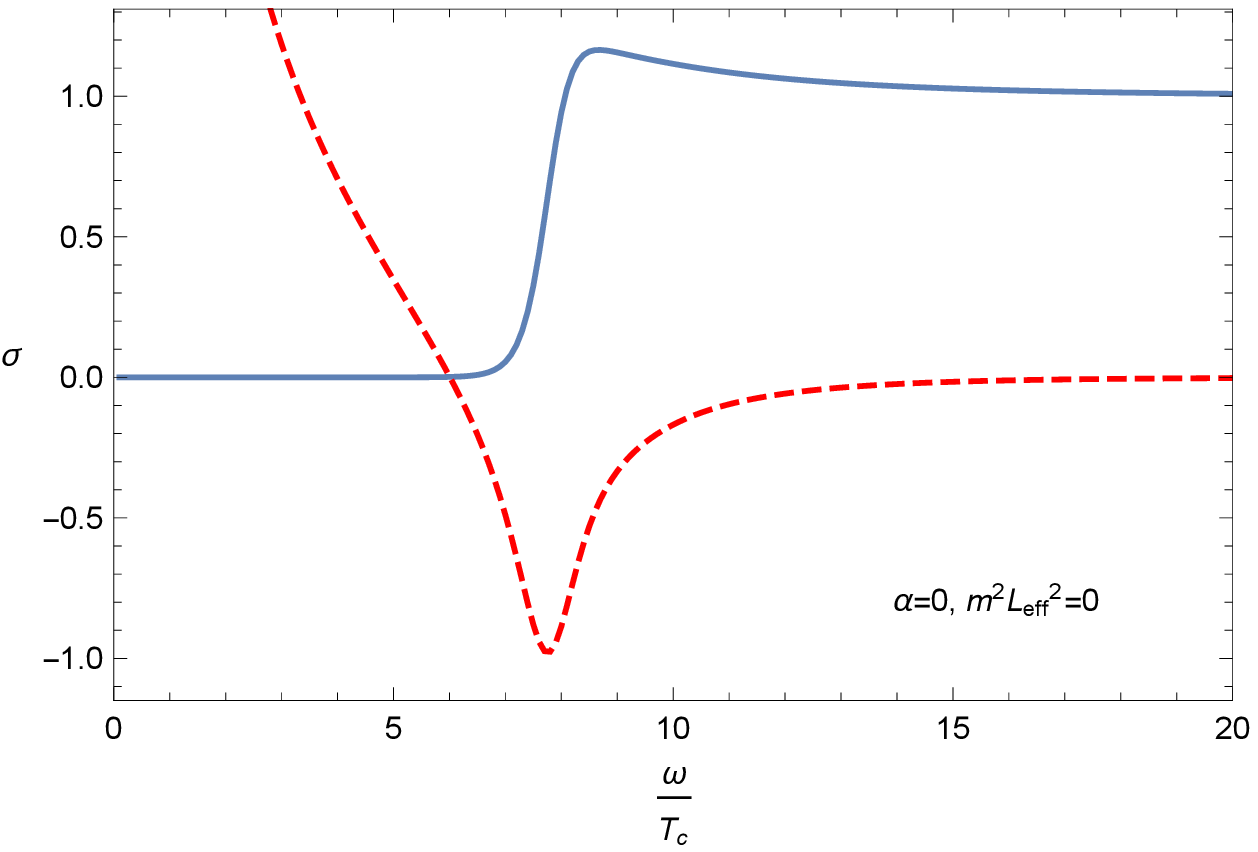}\hspace{0.2cm}%
\includegraphics[scale=0.425]{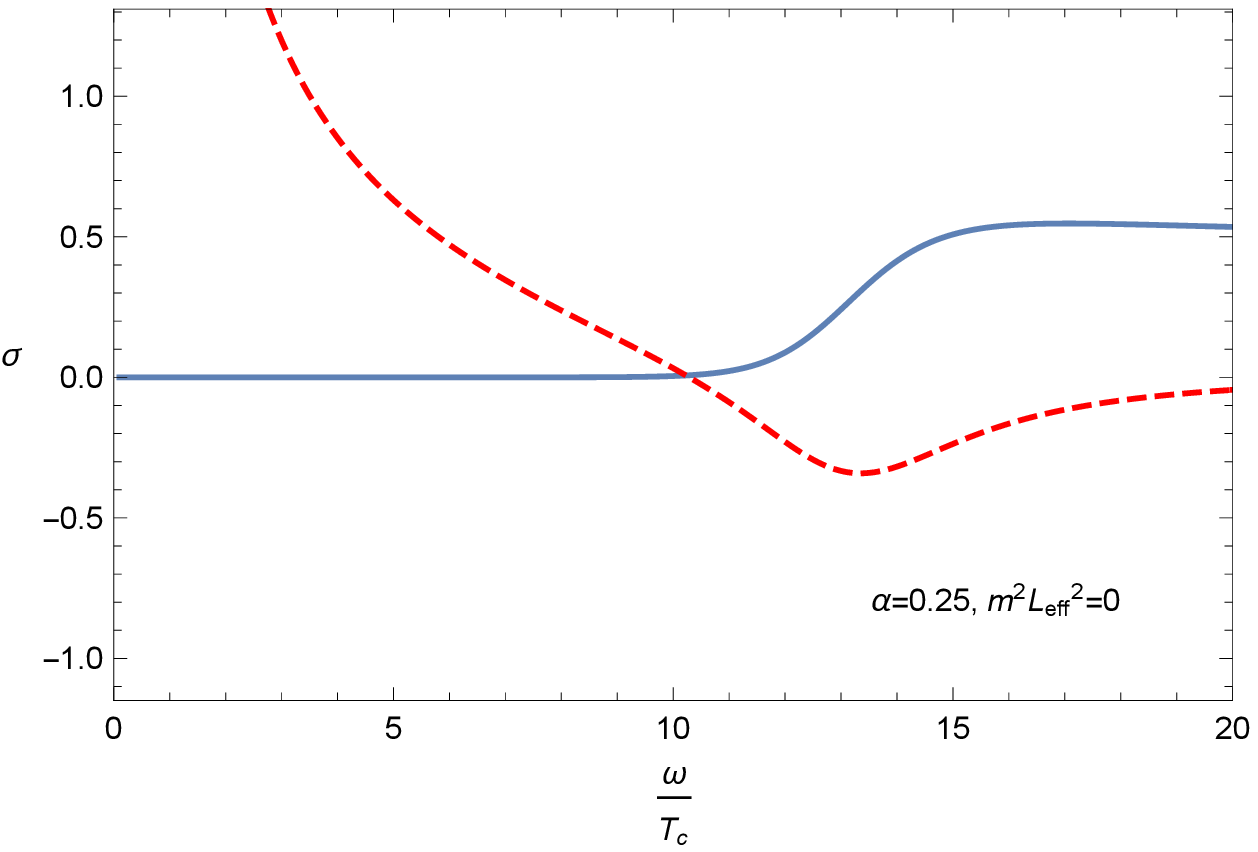}\\ \vspace{0.0cm}
\caption{\label{PWaveConductivity} (color online) Conductivity of the ($2+1$)-dimensional p-wave superconductors for different values of the Gauss-Bonnet parameter with the fixed vector field masses $m^{2}L_{\rm eff}^{2}=3/4$ and $0$. In each panel, the blue (solid) line and red (dashed) line represent the real part and imaginary part of the conductivity $\sigma(\omega)$ respectively.}
\end{figure}

In Fig. \ref{PWaveConductivity}, we show the frequency dependent conductivity of the ($2+1$)-dimensional p-wave superconductors in the Einstein-Gauss-Bonnet gravity for $\alpha=-0.1$, $0$ and $0.25$ with the fixed masses $m^{2}L_{\rm eff}^{2}=3/4$ and $0$ at temperatures $T/T_{c}\approx0.2$. Similar to the s-wave case in Fig. \ref{SWaveConductivity}, we find that the conductivity
develops a gap with the gap frequency $\omega_{g}$ and the larger deviation from the value $\omega_g/T_c\approx 8$ with the increase of $|\alpha|$. Thus, from Figs. \ref{SWaveConductivity} and \ref{PWaveConductivity}, our study implies that the higher curvature correction results in the larger deviation from the expected universal relation in the gap frequency $\omega_g/T_c\approx 8$ \cite{HorowitzPRD78} in both ($2+1$)-dimensional s-wave and p-wave superconductor models.

\section{Conclusions}

In this work, we first investigated the 4D neutral AdS black-hole solution in the consistent $D\rightarrow4$ Einstein-Gauss-Bonnet gravity proposed in \cite{AokiGMPLB}, which can also be obtained via the naive $D\rightarrow4$ limit of the higher-dimensional theory (by setting the charge $Q=0$) \cite{Fernandes,KonoplyaZhidenko} as well as the regularized 4D Einstein-Gauss-Bonnet gravity \cite{LuPang,HennigarKMP}. Then, we constructed the s-wave and p-wave holographic superconductors in this Gauss-Bonnet black hole background and analyzed the effect of the curvature correction on the ($2+1$)-dimensional superconductors, which may help to understand the influences of the $1/N$ or $1/\lambda$ (with $\lambda$ being the 't Hooft coupling) corrections on the holographic superconductor models. In the probe limit, we found that it is more appropriate to choose the mass of the field by selecting the value of $m^{2}L_{\rm eff}^{2}$, since this choice can disclose the correct consistent influence due to the Gauss-Bonnet parameter in various condensates for all dimensions. In the s-wave model, although the underlying mechanism remains mysterious, the effect of the curvature correction is more subtle: the critical temperature first decreases then increases as the Gauss-Bonnet parameter tends towards the Chern-Simons value in a scalar mass dependent fashion for the scalar operator $O_{+}$, but always decreases for the scalar operator $O_{-}$, which exhibits a very interesting and different feature when compared to the higher-dimensional Gauss-Bonnet superconductors. In the p-wave model, we noted that the critical temperature decreases as the Gauss-Bonnet parameter increases, which tells us that the higher curvature correction makes it harder for the vector hair to form in the full parameter space. On the other hand, for all cases considered here, we pointed out that the curvature correction will not modify the critical phenomena, i.e., the holographic superconductor phase transition belongs to the second order and the critical exponent of the system takes the mean-field value. Moreover, we observed that the higher curvature correction results in the larger deviation from the expected universal relation in the gap frequency $\omega_g/T_c\approx 8$ in both ($2+1$)-dimensional s-wave and p-wave superconductors, which is the same to the finding obtained in ($3+1$)-dimensions.

\begin{acknowledgments}

This work was supported by the National Natural Science Foundation of China under Grant Nos. 11775076, 11875025 and 11690034; Hunan Provincial Natural Science Foundation of China under Grant No. 2016JJ1012.

\end{acknowledgments}

\end{document}